\documentclass[11pt]{article}
\usepackage{graphicx,epic,amssymb,longtable,color,multibox,amsmath}
\definecolor{mygrey}{rgb}{0.5,0.5,0.5}

\newtheorem{prop}{Proposition}

\newcommand{\be}{\begin{eqnarray}}
\newcommand{\ee}{\end{eqnarray}}
\newcommand{\nn}{\nonumber}
\newcommand{\mf}{\mathfrak}
\newcommand{\mt}{\mathtt}
\newcommand{\bi}{\bibitem}

\def\ints{{\mathbb{Z}}}
\def\reals{{\mathbb{R}}}

\def\lra{\longleftrightarrow}
\def\lak{\mathfrak{k}}
\def\lag{\mathfrak{g}}

\def\fn{\footnotesize}

\def\pt{\partial_t}
\def\p{{\partial}}
\def\Pb{{\bar{P}}}
\def\Kb{{\bar{K}}}
\def\Qb{{\bar{Q}}}
\def\mb{{\bar{m}}}
\def\ab{{\bar{a}}}
\def\bb{{\bar{b}}}
\def\cb{{\bar{c}}}

\def\ch{{\hat{c}}}
\def\dh{{\hat{d}}}
\def\rh{{\hat{\rho}}}
\def\db{{\bar{\delta}}}
\def\eb{{\bar{e}}}

\def\H{{\mf{H}}}
\def\Vh{\hat{\mathtt{V}}}
\def\cP{{\mathcal{P}}}
\def\cG{{\mathcal{G}}}
\def\cQ{{\mathcal{Q}}}
\def\cV{{\mathcal{V}}}
\def\cJ{{\mathcal{J}}}
\def\cM{{\mathcal{M}}}
\def\cL{{\mathcal{L}}}
\def\O{\Omega}
\def\tO{\tilde{\Omega}}

\def\a{{\alpha}}
\def\b{{\beta}}
\def\g{{\gamma}}
\def\d{{\delta}}
\def\r{{\rho}}
\def\n{{\nu}}
\def\m{{\mu}}
\def\l{{\lambda}}
\def\s{{\sigma}}
\def\e{{\epsilon}}
\def\o{{\omega}}
\def\r{{\rho}}
\def\ve{\varepsilon}
\def\ra{\rightarrow}

\begin{document}

{\flushright AEI-2005-107\\hep-th/0506238\\[1cm]}

\begin{center}
{\LARGE {\bf  \Large \bf Gradient Representations and Affine
    Structures in $AE_n$}}\\[3cm]
{\bf Axel Kleinschmidt and Hermann Nicolai}\\
Max--Planck--Institut f\"ur Gravitationsphysik
(Albert--Einstein--Institut)\\
M\"uhlenberg 1, D-14476 Golm, Germany\\
{\tt axel.kleinschmidt,hermann.nicolai@aei.mpg.de}\\[30mm]
\end{center}

\begin{center}
{\bf \sc Abstract}\\[3mm]
\begin{tabular}{p{12cm}}
{\small We study the indefinite Kac--Moody algebras $AE_n$, arising in the
reduction of Einstein's theory from $(n+1)$ space-time dimensions
to one (time) dimension, and their distinguished maximal regular subalgebras
$A_{n-1}\equiv\mf{sl}_n$ and $A^{(1)}_{n-2}$. The interplay between
these two subalgebras is used, for $n=3$, to determine the commutation
relations of the `gradient generators' within $AE_3$. The low level
truncation of the geodesic $\s$-model over the coset space $AE_n/K(AE_n)$
is shown to map to a suitably truncated version of the $SL(n)/SO(n)$
non-linear $\s$-model resulting from the reduction Einstein's equations
in $(n+1)$ dimensions to (1+1) dimensions. A further truncation to
diagonal solutions can be exploited to define a one-to-one correspondence
between such solutions, and null geodesic trajectories on the
infinite-dimensional coset space $\H/K(\H)$, where $\H$ is the (extended)
Heisenberg group, and $K(\H)$ its maximal compact subgroup. We clarify
the relation between $\H$ and the corresponding subgroup of the Geroch
group.}
\end{tabular}
\end{center}

\newpage

\begin{section}{Introduction}

Infinite-dimensional symmetries in gravity were first discovered by
Geroch \cite{Geroch} in the context of (3+1)-dimensional general
relativity after dimensional reduction to (1+1) dimensions (see
\cite{Hoens,McCallum} for an introduction and further references).
The group structure was
later shown to be associated with the affine extension of Ehlers's
$SL(2,\reals)$ symmetry \cite{Julia,BrMa87}, and similar affinizations
of hidden symmetries were discovered for other (super-)gravity
theories \cite{Julia,BrMa87,Nicolai:1991tt}. Evidence for the emergence
of an even larger symmetry corresponding to the Kac--Moody theoretic
over-extension $AE_3$ of the original $SL(2,\reals)$ was presented in
\cite{DaHeJuNi01}, following earlier suggestions of \cite{Julia,Nic}.
For gravity in $(n+1)$ space-time dimensions, the conjectured symmetry
is $AE_n$, while for eleven-dimensional supergravity it is $E_{10}$,
which contains $AE_{10}$ as a subalgebra governing the gravitational
sector of this theory.

In \cite{DaHeNi02} this conjecture was re-examined using the insights
from the study of cosmological billiards (reviewed in \cite{DaHeNi03}).
Besides a remarkable dynamical match of a certain `geodesic'
one-dimensional $\s$-model with the gravity theory, the conjecture was
made that the Kac--Moody algebra\footnote{We often use the acronyms
`KMA' for `Kac--Moody Algebra' and `CSA' for `Cartan subalgebra'.}
allows for a re-emergence of the dependence on the coordinates along
which the theory had been reduced. This would entail a `dimensional
transmutation', in the sense that the evolution of the geometrical data
of a higher-dimensional theory, usually governed by a set of partial
differential equations, can be mapped to a {\em one-dimensional null-geodesic
motion on some infinite-dimensional coset space}. This conjecture was
based on the observation that the hyperbolic algebra $E_{10}$ contains 
a set of generators possessing the correct structure for higher order 
gradients in the suppressed directions. These, and their analogs in $AE_n$,
will be called `gradient generators' in the present paper. Further 
progress in the study of the one-dimensional $\s$-model is partly 
(but not only) hindered by the lack of known commutation relations. 
Although the irreducible representations appearing in level expansions 
of these algebras w.r.t. to their $\mf{sl}_n$ subalgebras can be determined 
rather efficiently on the computer \cite{NiFi03}, the commutators are much 
harder to obtain\footnote{See also remarks after (\ref{enfn}) to appreciate 
the challenge.}. Some  progress on this front was reported recently 
in \cite{Fi05} where an algorithm for computing commutation relations 
in a Borel subalgebra was outlined. Different aspects of the one-dimensional 
model were studied in \cite{KlNi04a,DaNi04,KlNi04b}.

In this paper we study the $\s$-model based on $AE_n$, extending
previous results of \cite{DaHeNi03}, in order to examine aspects of
the general picture explained above. Our focus is on $AE_n$ rather
than $E_{10}$, since the core difficulties which one encounters in
matching the one-dimensional $\s$-model and the higher dimensional
field equations, appear right away at levels $\pm 1$ in a graded
decomposition of $AE_n$ under its $\mf{sl}_n$ subalgebra. In
other words, the key problem of elevating the linear duality of
free spin-2 theories to the non-linear level must be faced already
in the first step, whereas for $E_{10}$, the difficulties become
only visible at $\ell=3$ and beyond, because the duality relating
the 3- and 6-form fields is still a {\em linear} one, modulo
metric factors, just like in Maxwell theory.

As we will show, the `gradient representations' are intimately linked
to the affine subalgebra $A_{n-2}^{(1)}$ of $AE_n$. Exploiting the
interplay of this affine subalgebra with the finite-dimensional
$A_{n-1}\equiv \mf{sl}_n$ subalgebra we are able to derive an infinite
new set of structure constants for $AE_3$. After restriction of the
$AE_n$ $\s$-model to the affine $A_{n-2}^{(1)}$ subsector we exhibit
a map between special solutions of the $\s$-model and solutions of
the gravitational field equations with $(n-1)$ commuting spacelike
Killing vectors, corresponding to a reduction from $(n+1)$ dimensions
to $(1+1)$ dimensions, but with a restricted space dependence.
These results are analogous to previous ones in
\cite{DaHeNi02,DaHeNi03,KlNi04a,DaNi04,KlNi04b}, but permit us to
to expose the remaining discrepancies in the simplest possible context.
In order to focus on these difficulties, we truncate the affine
$\s$-model further to a `Heisenberg coset model' $\H/K(\H)$, which
can be solved exactly, and whose gravitational counterpart corresponds
to diagonal metric solutions with two commuting (spacelike) Killing
vectors (known as `polarised Gowdy cosmologies' \cite{CIM90,KiRe98}).
We will then exhibit an explicit one-to-one correspondence between
these two models, by showing how the general initial data of the
Heisenberg $\s$-model and the null geodesic trajectories on $\H/K(\H)$
which they generate, can be mapped to a general space- and time-dependent
diagonal metric configuration satisfying Einstein's equation. In this way,
we are able to validate the `gradient hypothesis', and thereby the main
conjecture of \cite{DaHeNi02}, at least in this simplified context.

We also elucidate the relation between the standard action
of the (restricted) Geroch group on diagonal solutions, and the
action of the Heisenberg group $\H$ on the null geodesics. More
specifically, we will show that the action of these two groups
coincides on the domain where the Geroch group acts non-trivially.
Let us recall that only `half' of the Geroch group acts non-trivially
in the standard realisation \cite{Hoens,BrMa87}, whereas the other half
merely shifts the integration constants arising in the definition
of the higher order `dual potentials', and has no effect on the
physical metric. For this reason, the standard Geroch group affects
only part of the initial data (the full initial data of the off-diagonal 
degrees of freedom, but only the `initial coordinates' of the scale
factors). By contrast, the realization proposed here is such that 
Heisenberg subgroup of the Geroch group acts non-trivially on {\em all} 
initial data. 
%The $\s$-model realisation 
%of the full affine $A_{n-2}^{(1)}$ will exhibit similar features in 
%comparison with the standard implementation of the Geroch group.

Other attempts to generate space-time dependence through algebraic
constructions based on $E_{11}$ \cite{We01} and similar `very-extended'
algebras \cite{EnHoTaWe03} have been developed and discussed in
\cite{We03b,EnHo04a,KlWe04}. In \cite{We03b} this was achieved by
including a certain irreducible representation of the relevant
very-extended algebra containing translation generators and
(infinitely many) other generalized central charges, whereas in
\cite{EnHo04a} space-time is thought to occur through an auxiliary
parameter and gradient representation in restricted models. A key
difference between the present approach and \cite{We01} is that
the correspondence exhibited here works {\em only after the gauge
degrees of freedom on both sides have been eliminated by suitable
gauge choices} : on the $\s$-model side this implies the complete
elimination of the degrees of freedom associated with the maximal
compact subgroups, whereas on the gravitational side, it involves
not only gauge-fixing the vielbein (as discussed in section~5.1),
but also solving the canonical constraints. As a consequence, the
global symmetry relates solutions which are {\em physically distinct}.
By contrast, the proposal of \cite{We01,We03b} seeks a `covariant 
formulation'. This means, that the symmetry is actually much larger 
than the relevant very-extended Kac--Moody group, as it must contain 
general coordinate transformations (for instance, via the closure with 
the conformal group \cite{We00}) and other gauge transformations; 
hence, one must disentangle transformations relating gauge equivalent 
configurations from those which generate physically inequivalent
solutions. While it is not possible to really discriminate between the 
different proposals by analysing low level degrees of freedom,
the issue will be decided by whether and how the higher level fields
can be fitted into the scheme. The present paper is a first step in
this direction.

\begin{figure}[t]
\begin{center}
\scalebox{1}{
\begin{picture}(180,80)
\put(3,25){$\a_1$}
\put(43,25){$\a_2$}
\put(83,25){$\a_3$}
\put(10,40){\circle*{10}}
\thicklines \multiput(50,40)(40,0){2}{\color{mygrey}\circle*{10}}
\multiput(15,40)(40,0){2}{\line(1,0){30}}
\put(145,40){\color{mygrey}\circle*{10}}
\put(135,25){$\a_{n-1}$}
\multiput(95,40)(10,0){5}{\line(1,0){5}}
\put(100,80){\circle{10}}
\put(93,95){$\a_n$}
\put(54,44){\line(4,3){43}}
\put(104,76){\line(6,-5){38}}
\end{picture}}
\caption{\label{aendynk}\sl The Dynkin diagram of $AE_n$ with
labelling of the simple roots shown. Solid nodes (black and grey) belong to
$A_{n-1}\equiv\mathfrak{sl}_n$ which gets enhanced to
$\mathfrak{gl}_n$. Grey nodes mark the common `horizontal'
$A_{n-2}\equiv\mf{sl}_{n-1}$ subalgebra.}
\end{center}
\end{figure}
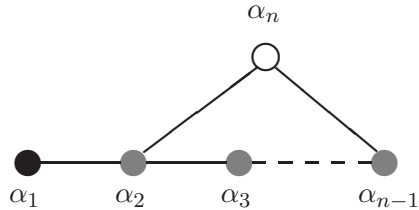

This article is structured as followed. First we define and
analyse the KMA $AE_n$ and its two distinguished maximal regular
subalgebras $A_{n-2}^{(1)}$ and $A_{n-1}$ and define the gradient
representations in this context in section \ref{aensubs}. In
section \ref{combaffin}, we compute infinitely many new structure
constants involving the gradient generators
in the special case of $AE_3$ by combining these two subalgebras.
The $\s$-model based on $AE_n$ is defined in section
\ref{sigmamod} where we study its restriction to $A_{n-2}^{(1)}$.
(The restriction to $A_{n-1}$ is studied in appendix
\ref{pformapp} where we also study gravity coupled to $p$-forms.)
The map from (parts of) the affine model to the two-dimensional
reduction of gravity is deduced in section \ref{2dsec}. Finally,
by using the Heisenberg model we examine the relation of the
$\s$-model to the Geroch group in section \ref{heissec}.
Appendix \ref{compapp} contains a proof of the fact that the compact
subalgebra of a (split) KMA is {\em not} a KMA.

\end{section}

\begin{section}{Distinguished maximal regular subalgebras of $AE_n$}
\label{aensubs}

\begin{subsection}{Definition of $AE_n$}

The indefinite Kac--Moody algebras $AE_n$ \footnote{We will designate by
$AE_n$ {\em both} the group {\em and} the Lie algebra, as it should be clear
from the context which is meant.} (for $n\geq 3$) are defined in the
usual way via the Chevalley--Serre presentation \cite{Ka90} associated
with the Dynkin diagram displayed in figure \ref{aendynk}. The simple
positive and negative generators are denoted by $e_a$ and $f_a$ respectively,
the Cartan subalgebra (CSA) generators by $h_a$, where $a=1,2,\ldots,n$.
We consider the algebra over $\reals$ in split real form. $AE_n$ is hyperbolic
for $n\le9$; this means that all Dynkin subdiagrams obtained by removing
one or more nodes are either affine or finite. We will also need to make
use of the Chevalley involution $\o$, defined on the simple generators by
\be\label{inv}
\o(e_a)=-f_a,\quad\quad \o(f_a)=-e_a,\quad\quad\o(h_a)=-h_a,
\ee
and the generalized transposition
\be\label{T}
x^T := -\o(x).
\ee
The `maximal compact' subalgebra $K(AE_n)$ of $AE_n$ is then
defined as the invariant algebra w.r.t. the involution $\o$, viz.
\be
\label{compsub}
K(AE_n) := \{ x\in AE_n \, | \, x = \o (x) \}
\ee
and consists of the `antisymmetric' Lie algebra elements in view of
(\ref{T}).

In this section we shall consider two distinguished maximal regular
subalgebras of
$AE_n$, namely $\mf{sl}_n$ and the affine $A_{n-2}^{(1)}$, obtained
by removing the nodes labeled $n$ and $1$, respectively. The first
generates the group of special linear transformations acting on the spatial
$n$-bein of gravity in $(n+1)$ space-time dimensions, and can be
enlarged to $\mf{gl}_n$ by inclusion of the CSA generator associated
with the white node in figure~\ref{aendynk}.
The second subalgebra corresponds to the
generalization of the Geroch group that is obtained after reduction
of pure gravity from $(n+1)$ dimensions to $(1+1)$ dimensions.
\end{subsection}

\begin{subsection}{$A_{n-1}\equiv\mf{sl}_n$ Subalgebra and level
decomposition of $AE_n$}
\label{slnsub}

The regular subalgebra of type $A_{n-1}\equiv\mf{sl}_n$ is generated
by considering only commutators of the simple generators associated
with nodes $1$ up to  $(n-1)$ in figure \ref{aendynk}. By including the
Cartan subalgebra element $h_n$ one can extend $\mf{sl}_n$ to $\mf{gl}_n$.
Its generators are denoted by $K^a{}_b$ ($a,b=1,2,\ldots,n$) and obey
the standard commutation relations
\be
\label{glncomm}
[K^a_{\ b}, K^c_{\ d}] = \d^c_b K^a_{\ d} -\d^a_d K^c_{\ b}.
\ee
For these generators, the transposition (\ref{T}) reduces to
$(K^T)^a_{\ b} = K^b_{\ a}$. (\ref{glncomm}) entails the following
identification of the $\mf{gl}_n$ elements with the Chevalley-Serre
generators of $AE_n$ for $i=1,2,\ldots,n-1$
\be\label{Chev1}
e_i= K^i{}_{i+1},\qquad f_i=K^{i+1}{}_i,\qquad
h_i=K^i{}_{i}-K^{i+1}{}_{i+1}.
\ee
Regularity of the $\mf{gl}_n$ subalgebra means that the standard
invariant bilinear form on $AE_n$, which is defined by
\be\label{bilinear0}
\langle e_i | f_j\rangle = \d_{ij}  \quad ; \qquad
\langle h_i | h_j\rangle = A_{ij}
\ee
coincides with the usual Cartan--Killing form on $\mf{gl}_n$ when
restricted to the latter subalgebra ($A_{ij}$ is the Cartan matrix of
$AE_n$, which is simply laced). Explicitly, the scalar product on
$\mf{gl}_n$ is given by
\be\label{bilinear1}
\langle K^a_{\ b} | K^c_{\ d}\rangle = \d^c_b\d^a_d -\d^a_b\d^c_d.
\ee
It is straightforward to check that this is indeed consistent
with (\ref{bilinear0}). Let us also express the trace 
$K=\sum_{a=1}^n K^a_{\ a}$ in $\mf{gl}_n$ in terms of the CSA generators
$h_a$; we have
\be
\label{traceiden}
K = -(n-1)h_1-(2n-2)h_2-(2n-3)h_3-\ldots-n\, h_n.
\ee
Solving for $h_n$ gives
\be
\label{hniden}
h_n=-K^1{}_1-K^2{}_2+K^n{}_n
=-K+K^3{}_3+K^4{}_4+\ldots+K^{n-1}{}_{n-1}+2K^n{}_n
\ee

The adjoint representation of $AE_n$ can be decomposed under the
(adjoint) action of $\mf{gl}_n$ into an infinite tower of $\mf{sl}_n$
representations. The $\mf{sl}_n$ level $\ell$, or simply the {\em level},
of a given representation counts the number of occurrences of the
simple root $\a_n$ in the corresponding $AE_n$ root $\a$, {\it i.e.}
\be
\a = \ell \a_n + \sum_{j=1}^{n-1} m^j \a_j
\ee
This level is left invariant by the action of $\mf{sl}_n$
and provides an elliptic slicing of the forward lightcone in the
$AE_n$ root lattice.
Level $\ell=0$ contains the adjoint of $\mf{gl}_n$. At level $\ell =1$
we have the representation
\be\label{l1}
\label{dualgravstart}
[0,1,0,0,\ldots,0,1] \lra E^{a_1\ldots a_{n-2},a_{n-1}}
\ee
associated with the Young tableau
\begin{center}
\scalebox{1}{\begin{picture}(50,85)
\multiframe(0,65)(0,0){1}(20,20){\fn $a_1$}
\multiframe(20,65)(0,0){1}(20,20){\fn $a_{n-1}$}
\multiframe(0,45)(0,0){1}(20,20){\fn $a_2$}
\multiframe(0,20)(0,0){1}(20,25){\fn $\vdots$}
\multiframe(0,0)(0,0){1}(20,20){\fn $a_{n-2}$}
\end{picture}}
\end{center}
This tensor is therefore antisymmetric in its first $(n-2)$ indices
$[a_1\dots a_{n-2}]$, and obeys
\be
\label{dualgravend}
E^{[a_1\ldots a_{n-2},a_{n-1}]} &=& 0.
\ee
Under Chevalley transposition we obtain
\be
F_{a_1\ldots a_{n-2},a_{n-1}} \equiv -\o(E^{a_1\ldots a_{n-2},a_{n-1}}).
\ee
Under $\mf{gl}_n$ the tensors $E^{a_1\ldots a_{n-2},a_{n-1}}$ and
$F_{a_1\ldots a_{n-2},a_{n-1}}$ transform contravariantly or
covariantly, as indicated by the position of indices, for instance
\be
[K^a{}_b , E^{c_1\ldots c_{n-2},c_{n-1}}] =
\d^{c_1}_b E^{a c_2\ldots c_{n-2},c_{n-1}} + \cdots +
\d^{c_{n-1}}_b E^{c_1\ldots c_{n-2},a}
\ee
Hence the level is counted by the operator $\frac1{n-1} K$.

The identification of the $AE_n$ Chevalley--Serre basis is completed
by relating the generators $e_n$ and $f_n$ to the level $\ell = \pm 1$
generators via
\be
e_n=E^{3\,4\ldots n,n},\quad\quad f_n=F_{3\,4\ldots n,n}.
\ee
The commutator $[(\ell\!=\! 1),(\ell\!=\! -1)]$ is best written using an
auxiliary (dummy) tensor $X_{a_1,\ldots a_{n-2},a_{n-1}}$ (with the same
Young symmetries as $F$) in the form
\be
\label{dualgravcomm}
&&\big[X_{a_1\ldots a_{n-2},a_{n-1}} E^{a_1\ldots a_{n-2},a_{n-1}},
    F_{b_1\ldots b_{n-2},b_{n-1}}\big]
= - (n-2)! \bigg(X_{b_1\ldots  b_{n-2},b_{n-1}} K\nn\\
&&\quad\quad\quad - X_{b_1\ldots b_{n-2},e} K^e{}_{b_{n-1}} - (n-2)
K^e{}_{[b_{n-2}} X_{b_1\ldots b_{n-3}]e,b_{n-1}}\bigg).
\ee
This is consistent with the normalisation\footnote{Symmetrization and
antisymmetrization is defined with `strength one' throughout this paper,
such as for instance in
$$
\d^{a_1a_2}_{b_1b_2} := \frac12 \d^{a_1}_{b_1}\d^{a_2}_{b_2} -\frac12
  \d^{a_1}_{b_2}\d^{a_2}_{b_1}\quad , \qquad
\db^{a_1a_2}_{b_1b_2} :=  \frac12 \d^{a_1}_{b_1}\d^{a_2}_{b_2} +\frac12
  \d^{a_1}_{b_2}\d^{a_2}_{b_1}.
$$
where a bar over $\delta$ denotes symmetrisation, and no bar means
antisymmetrisation.}
\be
&&\langle E^{a_1\ldots a_{n-2},{n-1}}| F_{b_1\ldots
  b_{n-2},b_{n-1}}\rangle\\
&&\quad=-\frac{n-2}{n-1} (n-2)! \bigg(\d^{a_1\ldots a_{n-2}}_{b_1\ldots
  b_{n-2}}\d^{a_{n-1}}_{b_{n-1}} + \d^{[a_1}_{b_{n-1}}\d^{a_2\ldots
    a_{n-2}]}_{[b_2\ldots b_{n-2}}\d^{a_{n-1}}_{b_1]}\bigg),\nn
\ee
corresponding to the standard normalisation for the Chevalley generators
\be\label{enfn}
\langle e_n | f_n\rangle =1.
\ee

All higher levels in the $\mf{sl}_n$ decomposition can be obtained by
taking multiple commutators of the level $\ell=1$ generator
$E^{a_1\ldots a_{n-2},a_{n-1}}$. For
$AE_3$, the decomposition into irreducible representations of $\mf{sl}_3$
is known up to level $\ell=56$ \cite{NiFi03}. Counting outer multiplicities,
the total number of $\mf{sl}_3$ representations up to that level is \cite{TF}
\be
\# \big(\mbox{representations for $\ell\leq 56$}\big) =
20\,994\,472\,770\,550\,672\,476\,591\,949\,725\,720
\ee
Consequently, the complete table of structure constants up to that
level will already contain more than $10^{62}$ entries! Hence, the
`gradient representations' that we will consider below constitute only a
tiny subsector (but not a subalgebra) of the full Lie algebra.

\end{subsection}

\begin{subsection}{Affine $A_{n-2}^{(1)}$ Subalgebra}
\label{affsubsec}

Figure \ref{aendynk} also shows that $AE_n$ has a regular
affine subalgebra $A_{n-2}^{(1)}\equiv \widehat{\mf{sl}_{n-1}}\oplus
\reals\ch\oplus\reals\dh$ generated
by nodes 2 up to  $n$ (the circular subdiagram in
figure~\ref{aendynk}).\footnote{We follow the convention that
  $\widehat{\mf{sl}_{n-1}}$ denotes the loop algebra based on
  $A_{n-2}\equiv\mf{sl}_{n-1}$. The non-twisted affine algebra
  $A_{n-2}^{(1)}$ is obtained by adjoining the central element
  $\ch$ and the derivation $\dh$ to the Cartan subalgebra, which
  then has dimension $n$ \cite{Ka90}. (The rank of $A_{n-2}^{(1)}$
  is $n-1$.)} We write
the corresponding {\em traceless} generators as $\Kb_{m}^\a{}_\b$ with
$m\in\ints$, and Greek indices $\a,\b,\dots \in \{2,\ldots,n\}$.
The generators which belong to both this affine subalgebra and the
$\mf{sl}_{n-1}$ constitute what we call the {\em horizontal algebra}
$\mathfrak{sl}_{n-1}$ corresponding to the nodes $2$ up to $n-1$,
with generators
\be
\label{idenfinaff}
\Kb_{0}^\a{}_\b=K^\a{}_\b-\frac1{n-1}\d^\a_\b K^\g{}_\g
\ee
in the $\mathfrak{gl}_n$  from above. The relevant nodes in
figure~\ref{aendynk} are marked in grey.

The affine commutation relations are
\be
\label{affinerels}
[\Kb_{m}^\a{}_\b,\Kb_{n}^\g{}_\d] = \d^\g_\b \Kb_{m+n}^\a{}_\d -
\d^\a_\d \Kb_{m+n}^\g{}_\b + m \d_{m,-n} \left(\d^\a_\d \d^\g_\b
-\frac1{n-1} \d^\a_\b\d^\g_\d\right)\, \ch.
\ee
The central element $\ch$ and the derivation $\dh$ required for the
$n$-dimensional CSA of $A_{n-2}^{(1)}$ can be identified in the
CSA of the rank $n$ KMA $AE_n$. Requiring the affine level counting
operator $\dh$ to obey
\be
[\dh,e_n]=-e_n,\quad, \qquad [\dh,e_i]=0 \quad(i=2,\ldots,n-1),
\quad\quad \langle\dh|\dh\rangle=0,
\ee
results in the following expressions for $\ch$ and $\dh$ in terms
of the $\mathfrak{gl}_n$ elements
\be
\label{canddiden}
\ch&=&\sum_{i=2}^n h_i = -K^1{}_1,\\
\dh &=& \frac{n-2}{2(n-1)} K^1{}_1 -\frac1{n-1} K^\a{}_\a,
\ee
such that
\be
\label{tracek}
K=K^a{}_a&=&K^1{}_1+K^\a{}_\a=-(n-1)\dh-\frac{n}2\ch,\\
&\Longrightarrow& K^\a{}_\a=-(n-1)\,\dh-\frac{n-2}2\,\ch.\nn
\ee
The CSA elements $\ch$ and $\dh$ are normalised according to
\be
\langle \ch|\dh\rangle =-1 \quad , \qquad
\langle \dh| \dh\rangle = \langle \ch | \ch \rangle = 0,
\ee
and the other non-vanishing inner product is
\be
\label{affip}
\big\langle \Kb_{m}^\a{}_\b|\Kb_{n}^\g{}_\d\big\rangle
= \d_{m,-n}\left(\d^\a_\d\d^\g_\b-\frac1{n-1}\d^\a_\b\d^\g_\d\right),
\ee
which is compatible with (\ref{bilinear0}).

Alternatively, the adjoint of $AE_n$ can be decomposed under the action
of $A_{n-2}^{(1)}$. In this case, the grading is labeled
by the {\em affine level}, which is equal to the coefficient
of the root $\a_1$ (now providing a parabolic slicing of the forward
lightcone in the root lattice). This decomposition was introduced
and studied in \cite{FeFr83} for $AE_3$, but we will not require
these results here.\footnote{Beyond the so-called basic representation
  on level one there are already infinitely many $A_{n-2}^{(1)}$
  representations on affine level
  two, which can be calculated from modularity and an appropriate
  implementation of the Serre relations \cite{FeFr83}. This is true
  not only for $AE_3$ but also for $AE_n$.}

\end{subsection}

\begin{subsection}{Horizontal $\mf{sl}_{n-1}$ and Gradient Representations}
\label{horgrad}

Writing the Kac--Moody algebra $AE_n$ as a graded representation of
its $\mf{sl}_n$ subalgebra is not independent of making use of the
affine $A_{n-2}^{(1)}$ subalgebra. This is due to the identification
of common generators in the horizontal algebra
$\mf{sl}_{n-1}\equiv A_{n-2} = A_{n-1} \cap A_{n-2}^{(1)}$.
We repeat that this is the $A_{n-2}$ subalgebra of $AE_n$ generated by
the $n-2$ (grey)
nodes $2$ up to $n-1$  in diagram \ref{aendynk}, with the generators
given in eq.~(\ref{idenfinaff}). In particular, the affine generator
$\Kb_{m}^\a{}_\b$ on affine level $m$, transforming in the adjoint
of the horizontal $\mf{sl}_{n-1}$, also transforms under
$\mf{sl}_n\supset \mf{sl}_{n-1}$, and it is natural to ask to which
$\mf{sl}_n$ representation it belongs. This is most easily determined
by considering the lowest weight vector and its Dynkin labels with
respect to $\mf{sl}_n$. The lowest weight vector lies in the root
space of the level-$m$ $AE_n$ root (assuming $m\geq 1$ from now on)
\be
\a = m\a_n + (m-1) \sum_{j=2}^{n-1} \a_j
\ee
corresponding to $\mf{sl}_n$ Dynkin labels $[m-1,1,0,\ldots,0,1]$. We
see that this is consistent with the common horizontal
$\mf{sl}_{n-1}$: In terms of the Dynkin
labels restricting to the horizontal $A_{n-2}$ of
$A_{n-2}^{(1)}$ means dropping the first entry of
$[m-1,1,0,\ldots,0,1]$ and the
resulting representation is indeed simply the adjoint $[1,0,\ldots,0,1]$
of the horizontal $\mf{sl}_{n-1}$ as anticipated. The generator
corresponding to the $[m-1,1,0,\ldots,0,1]$ representation of
$\mf{sl}_n$ is therefore associated with the following Young tableau
\begin{center}
\scalebox{1}{\begin{picture}(100,105)\label{lm1}
\multiframe(0,0)(0,0){1}(20,20){}
\multiframe(0,20)(0,0){1}(20,20){}
\multiframe(0,40)(0,0){1}(20,25){\fn $\vdots$}
\multiframe(0,65)(0,0){1}(20,20){}
\multiframe(0,85)(0,0){1}(20,20){}
\multiframe(20,0)(0,0){1}(25,20){\fn $\cdots$}
\multiframe(20,20)(0,0){1}(25,20){\fn $\cdots$}
\multiframe(20,40)(0,0){1}(25,25){}
\multiframe(20,65)(0,0){1}(25,20){\fn $\cdots$}
\multiframe(20,85)(0,0){1}(25,20){\fn $\cdots$}
\multiframe(45,0)(0,0){1}(20,20){}
\multiframe(45,20)(0,0){1}(20,20){}
\multiframe(45,40)(0,0){1}(20,25){\fn $\vdots$}
\multiframe(45,65)(0,0){1}(20,20){}
\multiframe(45,85)(0,0){1}(20,20){}
\multiframe(65,85)(0,0){1}(20,20){\fn $ c_1 $}
\multiframe(85,85)(0,0){1}(20,20){\fn $c_{n-1}$}
\multiframe(65,65)(0,0){1}(20,20){\fn $c_2$}
\multiframe(65,40)(0,0){1}(20,25){\fn $\vdots$}
\multiframe(65,20)(0,0){1}(20,20){\fn $c_{n-2}$}
\end{picture}}
\end{center}
This level-$m$ element of $AE_n$ thus has $(m-1)$ symmetric sets of
$(n-1)$ antisymmetric indices and then $(n-1)$ indices with the
structure of the representation on $\mf{sl}_n$ level $\ell=1$,
cf. (\ref{dualgravstart})--(\ref{dualgravend}). Because this is rather
cumbersome to write out, it is convenient to treat this tensor as
a representation of  $\mf{sl}_n$, and to dualise the sets of antisymmetric
indices by means of the $\mf{sl}_n$ $\ve$-symbol. The result is an
$\mf{sl}_n$ tensor with $(m-1)$ {\em lower} indices
\be
\label{gradgens}
E_{b_1\ldots b_{m-1}}{}^{c_1\ldots c_{n-2},c_{n-1}}
\ee
which is {\em symmetric} in these indices, while the upper indices
belong to the level-one representation (\ref{l1}). There is a similar
expression for the transposed $F$ generator. We stress that (\ref{lm1})
and (\ref{gradgens}) are equivalent {\em only} as representations of
$\mf{sl}_n$, but not $\mf{gl}_n$, as they carry different $\mf{gl}_1$
charges. The commutation relations must therefore be amended by a
correction term for the transformation under the trace $K$ in order to
account for this charge. We will come back to this point below for $AE_3$.

As noted in \cite{DaHeNi02} the infinite tower of representations
(\ref{gradgens}) has precisely the right structure corresponding to the
{\em spatial gradients} of the low level fields, such that each index $b_i$
becomes associated with a gradient operator $\partial/\partial x^{b_i}$,
and the irreducibility condition of the above Young tableau
translates to vanishing divergence. More specifically, for $AE_n$, the
relevant degree of freedom is the {\em dual graviton}, corresponding to
the above tableau with only the right two columns as in
(\ref{dualgravstart}), and appears already
at level $\ell =1$ (for $E_{10}$ it appears only at level $\ell
=3$).\footnote{The dual graviton representation has been discussed
  previously in
  \cite{Curtright:1980yk,Obers:1998fb,Hull:2001iu,DaHeNi02,Bekaert:2002uh,West:2002jj,Nieto:1999pn,Henneaux:2004jw}.}
For this reason, we will refer to these representations as
{\em gradient representations}. Let us emphasize that, except for
the lowest levels, this is so far only a kinematical
correspondence.\footnote{We note that there are generators
resembling a $k$-th spatial derivative (on level $\ell+k$) of any
given generator (on level $\ell$). These will generally occur with
exponentially increasing outer multiplicity as they belong to
a standard lowest weight representation the affine algebra. The gradient
representations (\ref{gradgens}) are distinguished by being
generated from the adjoint representation of the affine algebra and
hence have constant outer multiplicity, equal to $1$.}

The other main feature of the gradient representations is that they
contain the affine generators. More specifically, the latter are
obtained by performing a `dimensional reduction', restricting the
gradient indices ({\it i.e.} the lower indices in (\ref{gradgens}))
to the single value $b_1 = \cdots = b_{m-1}= 1$, and the upper indices to the
remaining values $\a,\b, \dots = 2,\dots, n$. This truncation
corresponds precisely to a dimensional reduction of Einstein's
theory from $(n+1)$ to (1+1) spacetime dimensions, with spacelike
Killing vectors $\partial/\partial x^2 , \dots, \partial/\partial x^n$,
and $x\equiv x^1$ as the surviving spatial coordinate. The precise
identification of $\Kb_{m}^\a{}_\b$ (for $m>0$) as part of the
$\mf{sl}_n$ tensor (\ref{gradgens}) is
\be\label{Kh1}
\Kb_{m}^\a{}_\b = \frac1{(n-2)!}\ve_{\b\g_1\ldots\g_{n-2}}
E_{\scriptsize \underbrace{ 1\ldots
    1}_{(m-1)}}{}^{\g_1\ldots\g_{n-2},\a},
\ee
where the indices $\a,\b,\dots = 2,\dots ,n$ have been restricted to
$\mf{sl}_{n-1}$.
For affine level $m<0$ the identification is
\be\label{Kh2}
\Kb_{m}^\a{}_\b = \frac1{(n-2)!}\ve^{\a\g_1\ldots\g_{n-2}} F^{\scriptsize
  \overbrace{1\ldots
    1}^{(m-1)}}{}_{\g_1\ldots\g_{n-2},\b}.
\ee
The correctness of these identifications is easy to check at levels
$|m|\le 1$.

One can now exploit the interplay between $\mf{sl}_n$ and
$A_{n-2}^{(1)}$ to determine the complete commutators
of the `gradient generators' modulo non-gradient representations.
This is achieved by writing down the general ansatz for a commutator
in $\mf{sl}_n$ covariant form, and then restricting to the horizontal
$\mf{sl}_{n-1}$ and using the the affine relations (\ref{affinerels}).
In the next section, we exemplify this procedure for $AE_3$.

\end{subsection}

\end{section}

\begin{section}{$AE_3$ and its gradient commutators via $\mf{sl}_3$ and
$A_1^{(1)}$}
\label{combaffin}
\setcounter{equation}{0}

\begin{subsection}{$AE_3$ in $\mf{sl}_3$ form}

The Dynkin diagram of $AE_3$ is displayed in figure \ref{ae3dynk}.
Before computing the contribution of the gradient generators to
commutators for $AE_3$ we give the explicit formul\ae\ for the
$\mf{sl}_3$ generators and the appropriate low $\mf{sl}_3$ level
commutators of $AE_3$.

\begin{figure}[t]
\begin{center}
\scalebox{1}{
\begin{picture}(100,40)
\put(10,25){$\a_1$} \put(50,25){$\a_2$} \put(90,25){$\a_3$}
\thicklines \put(10,40){\circle*{10}}
\put(50,40){\color{mygrey}\circle*{10}}
\multiput(15,40)(40,0){1}{\line(1,0){30}} \put(90,40){\circle{10}}
\multiput(50,35)(0,10){2}{\line(1,0){40}}
\end{picture}}
\caption{\label{ae3dynk}\sl The Dynkin diagram of $AE_3$ with
labelling of the simple roots. Solid nodes belong to $\mf{sl}_3$
which gets enhanced to $\mf{gl}_3$. The grey node marks the
horizontal $A_1\equiv\mf{sl}_2$.}
\end{center}
\end{figure}
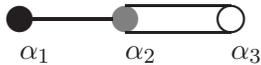

In this case, the general identifications (\ref{traceiden}) and
(\ref{hniden}) read
\be
h_3=-K+2K^3_{\ 3},\quad\quad
K=-2h_1-4h_2-3h_3.
\ee
The $\mf{gl}_3$ commutation relations and
inner products are identical in form to (\ref{glncomm}) but with
index ranges now restricted to $a=1,2,3$. The decomposition of
$AE_3$ under its $\mf{sl}_3$ subalgebra results in 
table~\ref{sl3cont}, containing levels $\ell=1,\ldots,5$.

\begin{table}[t]
{\center
\begin{tabular}{|c|c|c|c|c|c|c|}
\hline
$\ell$&$[p_1,p_2]$&$\a$&$\a^2$&$\textrm{mult}\a$&$\m$&Gradient\\
\hline
\hline
1&[0,2]&(0,0,1)&2&1&1&*\\
2&[1,2]&(0,1,2)&2&1&1&*\\
3&[2,2]&(0,2,3)&2&1&1&*\\
&[1,1]&(1,3,3)&-4&3&1&\\
4&[3,2]&(0,3,4)&2&1&1&*\\
&[2,1]&(1,4,4)&-6&5&2&\\
&[1,0]&(2,5,4)&-10&11&1&\\
&[0,2]&(2,4,4)&-8&7&1&\\
&[1,3]&(1,3,4)&-2&2&1&\\
5&[4,2]&(0,4,5)&2&1&1&*\\
&[3,1]&(1,5,5)&-8&7&3&\\
&[2,0]&(2,6,5)&-14&22&3&\\
&[0,1]&(3,6,5)&-16&30&2&\\
&[0,4]&(2,4,5)&-6&5&2&\\
&[1,2]&(2,5,5)&-12&15&4&\\
&[2,3]&(1,4,5)&-4&3&2&\\
\hline
\end{tabular}
\caption{\label{sl3cont}\sl Adjoint of $AE_3$ decomposed under
  regular $\mf{sl}_3$ subalgebra. The charge of an element $X$
  on level $\ell$ under the trace $K=K^a{}_a$ of $\mf{gl}_3$ is
  $[K,X]=2\ell X$ as explained in the text.}}
\end{table}

Generators belonging to the class of gradient generators have been
marked with an asterisk. Our notation for representations of
$\mf{gl}_3$ is the one inherited from $\mf{sl}_3$. The general
$\ell=1$ generator (\ref{dualgravstart}) reduces to a generator of the
type
\be
E^{b_1b_2},
\ee
where we have dropped the separating comma. The Young irreducibility
constraint (\ref{dualgravend}) now implies that this tensor is symmetric,
\be
\label{dualgrav3sym}
E^{b_1b_2}=E^{b_2b_1}.
\ee
For $\mf{sl}_3$ one can write any representation with Dynkin labels
$[p,q]$ as a tensor with $p$ lower and $q$ upper indices, after
lowering the $p$ pairs of antisymmetric indices with $\ve$ of
$\mf{sl}_3$. The irreducibility constraints are then symmetry in lower
and upper indices separately and vanishing of any contraction.
The totally antisymmetric tensor $\ve_{abc}$ used for lowering the
indices, however, is {\em not} invariant under $\mf{gl}_3$ as can
be checked easily. That means that while the $\ell=2$ tensor with
natural indices $E^{c_1c_2,b_1b_2}$ transforms under $\mf{gl}_3$
as the indices suggest, {\em i.e.}
\be
[K^a{}_b, E^{c_1c_2,b_1b_2}] = \d^{c_1}_b E^{ac_2,b_1b_2} +
  \d^{c_2}_b E^{c_1a,b_1b_2} + \d^{b_1}_b E^{c_1c_2,ab_2} +
  \d^{b_2}_b E^{c_1c_2,b_1a},
\ee
the corresponding tensor with lowered indices does {\em not}
transform in the obvious way. Rather one has to add a compensating
term for the transformation under the trace $K=K^a{}_a$. Generally,
\be\label{trmlower}
[K^c{}_d, E_{a_1\ldots a_k}{}^{b_1b_2}] = - k \d^c_{(a_1} E_{a_2\ldots
  a_k)d}{}^{b_1b_2}
  + 2\d^{(b_1}_d E_{a_1\ldots a_k}{}^{b_2)c}
  +k \d^c_d E_{a_1\ldots a_k}{}^{b_1b_2}.
\ee
The last term is the necessary correction: The charge of the
original tensor on level $k+1$ with $2k+2$ upper indices under the
trace $K$ is $2k+2$, and this charge has to be maintained. Similar
remarks apply to the Chevalley transposed generators
\be
F^{a_1\ldots a_p}{}_{b_1\ldots b_q}
  =(E_{a_1\ldots a_p}{}^{b_1\ldots b_q})^T.
\ee

We now present the commutation of the $AE_3$ generators
for levels $|\ell|\le 2$ in $\mf{sl}_3$ form. These can be deduced
from table \ref{sl3cont} after the generators have been normalised in
the standard bilinear form.\footnote{In general, there remain signs
  $\pm   1$ which need to be fixed consistently. Up to $|\ell|\le 2$
  our choice is consistent.} For $\ell=1,2$ we demand
\be
\langle E^{b_1b_2}_{\ }| F_{d_1d_2}^{\ }\rangle &=&
  \db^{b_1b_2}_{d_1d_2}:= \d^{(b_1}_{d_1}\d^{b_2)}_{d_2},\\
\langle E_a{}^{b_1b_2}| F^c{}_{d_1d_2}\rangle &=&
  P_a{}^{b_1b_2}|^c{}_{d_1d_2} :=
  \d^c_a \db^{b_1b_2}_{d_1d_2}
  -\frac12 \d^{(b_1}_a \d^{b_2)}_{(d_1}\d^c_{d_2)}.
\ee
The commutation relations are then
\be
[E^{b_1b_2},F_{d_1d_2}] &=& -\db^{b_1b_2}_{d_1d_2} K
  +2\d^{(b_1}_{(d_1}K^{b_2)}_{\ d_2)},\\
{}[E^{b_1b_2},E^{c_1c_2}] &=& 2 \ve^{b_1c_1a} E_a{}^{b_2c_2},\\
{}[F_{d_1d_2},F_{c_1c_2}] &=& -2 \ve_{d_1c_1a} F^a{}_{d_2c_2},\\
{}[E_a{}^{b_1b_2},F_{d_1d_2}]&=& 2\ve_{ae(d_1}
   \d^{(b_1}_{d_2)} E^{b_2)e},\\
{}[F^c{}_{d_1d_2},E^{b_1b_2}]&=& 2\ve^{ce(b_1}
   \d_{(d_1}^{b_2)} F_{d_2)e},\\
{}[E_a{}^{b_1b_2}, F^c{}_{d_1d_2}] &=&
   -P_a{}^{b_1b_2}|^c{}_{d_1d_2} K
   + 2 \d^c_a \d^{(b_1}_{(d_2}K^{b_2)}_{\ d_2)}\nn\\
&&\quad -\db^{b_1b_2}_{d_1d_2} K^c_{\ a}
   -\frac12 \d^{(b_1}_a K^{b_2)}{}_{(d_1}\d^c_{d_2)}.
\ee
Here, symmetrisation over lower or upper non-contracted indices is implicit.

\end{subsection}

\begin{subsection}{$A_1^{(1)}$ and gradient commutators}

The identification of the affine $A_1^{(1)}\subset AE_3$ subalgebra is
achieved through (cf. (\ref{canddiden}))
\be
\ch &=& h_2+h_3=-K^1{}_1 \;\; ,\quad  \dh=h_1+\frac54 h_2+\frac34
h_3=\frac14 K^1{}_1-\frac12 ( K^2{}_2+K^3{}_3)\nn\\
K&=&K^a{}_a=K^1{}_1+K^\g{}_\g=-2\dh-\frac32\ch,\quad
\Longrightarrow K^\g{}_\g=-2\dh-\frac12\ch.
\ee
The conventional choice for $AE_3$ would be to take node $3$ to be the
horizontal node instead of $2$ as we did, since this is more natural
from the extension point of view. This choice would correspond to the
Ehlers $SL(2,\reals)$, and it is also the choice taken by \cite{FeFr83}.
However, it does not reduce to the desired horizontal
$\mf{sl}_2\subset\mf{sl}_3$ subalgebra, and for this reason we here
make another choice, corresponding to the `Matzner--Misner' $SL(2,\reals)$
in the physical interpretation. The gradient generators (\ref{gradgens})
on level $\ell$ have Dynkin labels $[\ell-1,2]$ and after lowering the
sets of antisymmetric indices take the form
\be
E_{a_1\ldots a_{\ell-1}}{}^{b_1b_2}.
\ee

We now determine the contribution of the gradient generators
to $AE_3$ commutators by writing down for each commutator
the most general $\mf{sl}_3$ covariant ansatz compatible with the
level decomposition, and then matching this ansatz with the affine
commutation relations (\ref{affinerels}). Let us first note down
the projector onto a representation with Dynkin labels $[\ell,2]$:
\be
\label{projell}
P_{a_1\ldots a_\ell}{}^{b_1b_2}|^{c_1\ldots c_\ell}{}_{d_1d_2} &=&
 \db_{a_1\ldots a_\ell}^{c_1\ldots c_\ell} \db^{b_1b_2}_{d_1d_2}
 -\frac{2\ell}{3+\ell} \db^{c_1\ldots c_{\ell-1}}_{a_1\ldots
 a_{\ell-1}}\d^{(b_1}_{a_\ell}
 \d^{b_2)}_{(d_1}\d^{c_\ell}_{d_2)}\\
&&\quad +\frac{\ell(\ell-1)}{(3+\ell)(2+\ell)} \db^{c_1\ldots
 c_{\ell-2}}_{a_1\ldots a_{\ell-2}}
 \db^{b_1b_2}_{a_{\ell-1}a_\ell}
 \db^{c_{\ell-1}c_\ell}_{d_1d_2},\nn
\ee
where the right hand side also has to be symmetrized over the $a$
and $c$ indices. Here, $\db$ denotes the strength one symmetrizing projector.
The second and third term are only
present when $\ell\ge 1$ or $\ell\ge 2$ respectively. The
projector has the unusual property that it also serves as a
projector on the representation $[2,\ell]$ when acting from the
right.\footnote{That this is not naturally so can be seen, for
  instance, from considering the projector on the Riemann tensor
  symmetries.}

The restriction to those elements in $E_{a_1\ldots
a_{\ell-1}}{}^{b_1b_2}$ belonging strictly to the affine subalgebra
gives an identification of $\Kb_{\ell}^\a{}_\b$ as explained above.
Here, the identification of the affine subalgebra within $AE_3$ for
all levels $\ell>0$ reads
\be
\label{iden1}
\Kb_{\ell}^\a{}_\b = \ve_{\b\g}E_{\scriptsize \underbrace{1\ldots
    1}_{(\ell-1)}}{}^{\g\a}
\ee
and for $-\ell<0$
\be
\Kb_{-\ell}^\a{}_\b = \ve^{\a\g} F^{\scriptsize \overbrace{1\ldots
    1}^{(\ell-1)}}{}_{\g\b}.
\ee
These formulas specialize (\ref{Kh1}) and (\ref{Kh2}) to the case $n=3$.
Note that we use Euclidean conventions resulting in $\ve^{23}=\ve_{23}=1$.
It is easy to check that indeed ($\ell\geq 1$)
\be
\big[\dh, E_{\scriptsize \underbrace{1\ldots
    1}_{(\ell-1)}}{}^{\a\b}\big]
= - \ell E_{\scriptsize \underbrace{1\ldots 1}_{(\ell-1)}}{}^{\a\b}
\ee
and
\be
\big[\ch, E_{\scriptsize \underbrace{1\ldots
    1}_{(\ell-1)}}{}^{\a\b}\big] =0,
\ee
Here the correction term in (\ref{trmlower}) is essential in
order to obtain a vanishing result.

Upon restriction to the affine subalgebra the projectors
(\ref{projell}) simplify to
\be
P_{\scriptsize \underbrace{1\ldots 1}_{\ell}}{}^{\b_1\b_2}|^{c_1\ldots
  c_{\ell-1}}{}_{d_1d_2} = \db^{c_1\ldots c_{\ell}}_{1\ldots 1}
\db^{\b_1\b_2}_{d_1d_2}
\ee
for positive $\ell$ and for negative $\ell$ accordingly. The
identification (\ref{iden1}) together with the affine relations
(\ref{affinerels}) then determines the commutators of the gradient
representations. Note that the identification also fixes the
normalisation of the gradient tensor in the $AE_3$ invariant form to
be
\be
\langle E_{a_1\ldots a_\ell}{}^{b_1b_2}| F^{c_1\ldots
  c_k}{}_{d_1d_2}\rangle =\left\{\begin{array}{cc} P_{a_1\ldots
  a_\ell}{}^{b_1b_2}|^{c_1\ldots c_\ell}{}_{d_1d_2}&\textrm{for
  }\,\ell=k\\
0&\textrm{for }\, \ell\ne k \end{array}\right.,
\ee
by using eq.~(\ref{affip}).

The commutator of two $E$ generators on levels $\ell$ and $k$ is
then
\be
\label{eecomm}
[E_{a_1\ldots a_{\ell-1}}{}^{b_1b_2},E_{c_1\ldots c_{k-1}}{}^{d_1d_2}]
&=&  -E_{a_1\dots a_{\ell-1}c_1\ldots
  c_{k-1}e}{}^{d_1(b_1} \ve^{b_2)d_2e}\\
&& -  E_{a_1\dots a_{\ell-1}c_1\ldots
  c_{k-1}e}{}^{d_2(b_1} \ve^{b_2)d_1e}+\ldots.\nn
\ee
The dots on the right hand side indicate possible other $AE_3$
generators appearing in this commutators which will however vanish
upon restriction to the affine subalgebra $A_1^{(1)}$.
The term on the right is fixed up to normalisation from the tensor
symmetries and the normalisation is fixed from the affine
subalgebra. A similar expression holds for $[F,F]$.

For $\ell>k$ we obtain
\be
\label{efcomm}
&&[E_{a_1\ldots a_{\ell}}{}^{b_1b_2},F^{c_1\ldots c_{k}}{}_{d_1d_2}]
= -2 P_{a_1\dots a_\ell}{}^{b_1b_2}|^{e_1\ldots e_\ell}{}_{f_1f_2}
 P_{g_1\ldots g_k}{}^{h_1h_2}|^{c_1\ldots  c_k}{}_{d_1d_2} \nn\\
&&\quad\quad\cdot \ve_{e_{\ell}h_1i}\d_{h_2}^{f_1}\db_{e_1\ldots
 e_{k}}^{g_1\ldots g_{k}}
 E_{e_{k+1}\dots e_{\ell-1}}{}^{f_2i}+\ldots.
\ee

Finally, level $\ell+1>0$ commutes with level $-\ell-1$ into the
adjoint of $\mf{gl}_3$
\be
&&[E_{a_1\ldots a_\ell}{}^{b_1b_2},F^{c_1\ldots c_\ell}{}_{d_1d_2}]
 = P_{a_1\ldots a_\ell}{}^{b_1b_2}|^{e_1\ldots
 e_\ell}{}_{f_1f_2} P_{g_1\ldots g_\ell}{}^{h_1h_2}|^{c_1\ldots
 c_\ell}{}_{d_1d_2}\\
&&\,\,\cdot\bigg(\d^{g_1}_{e_1}\cdots\d^{g_\ell}_{e_\ell}\d^{f_1}_{h_1}
 \d^{f_2}_{h_2} K
 -2\d^{g_1}_{e_1}\cdots\d^{g_\ell}_{e_\ell}\d^{f_1}_{h_1}
 K^{f_2}{}_{h_2} +\ell \d^{f_1}_{h_1}\d^{f_2}_{h_2}
 \d^{g_1}_{e_1}\cdots\d^{g_{\ell-1}}_{e_{\ell-1}} K^{g_\ell}{}_{e_\ell}
\bigg).\nn
\ee
This commutator is exact since there are no other Lie algebra elements
besides those of $\mf{gl}_3$ at level $\ell=0$. The omitted terms
(=dots) in the other $AE_3$ commutators are necessary for the Jacobi
identities. This phenomenon first occurs at level $\ell=3$, since this
is the first time a non-gradient representation is present (cf. table
\ref{sl3cont}). Checking the Jacobi identities using solely the gradient
generators shows a violation which is resolved when one also takes into
account the $\ell=3$ representation with Dynkin labels $[1,1]$.

\end{subsection}

\end{section}

\begin{section}{Affine truncation of the $AE_n/K(AE_n)$ $\s$-model}
\label{sigmamod}
\setcounter{equation}{0}

Having clarified the embeddings of the subgroups and having determined
the gradient commutators, we next turn to the one-dimensional `geodesic'
$\s$-model over the coset space $AE_n/K(AE_n)$. This model is
governed by the Lagrangian \cite{DaHeNi03}
\be\label{Lag}
L=\frac1{2n}\langle \cP| \cP\rangle,
\ee
where all quantities depend only on the affine parameter (`time') $t$,
and $n(t)$ is the lapse function required for reparametrisation
invariance $t\rightarrow \tilde{t}(t)$. The quantity $\cP(t)$ and the
$K(AE_n)$ gauge connection $\cQ(t)$ are determined from the Cartan form
\be\label{QP}
\pt\cV\cV^{-1}=\cQ+\cP
\ee
with $\cV(t) \in AE_n/K(AE_n)$. By construction, $\cQ$ belongs to the
compact subalgebra $K(AE_n)$ (fixed by the Chevalley involution, cf.
eq.~(\ref{compsub})), and $\cP$ to its complement in $AE_n$. With
these definitions and notations, the equations of motion assume the
simple looking form \cite{DaNi04}
\be
\label{eom}
n \pt(n^{-1}\cP) = [\cQ,\cP],
\ee
To get these equations into the standard second order form, we would
have to choose coordinates on the $AE_n/K(AE_n)$ coset manifold and to
substitute them into (\ref{QP}), but we will skip this step here. Although
these objects are highly formal constructs at this point, readers need
not worry about our lack of knowledge of what the `group' $AE_3$ really
is, because the truncated level expansion provides an algorithm which
is such that all operations take place in the Lie algebra, and involve
only a finite number of steps if one truncates at finite level. As we
will show these steps remain well-defined for the {\em affine truncation}
where one retains an infinite number of Lie algebra elements.

Varying the lapse $n$, we get the (Hamiltonian) constraint
\be
\langle\cP | \cP\rangle = 0
\ee
Connsequently, the `trajectory' described by (\ref{eom}) on the
coset manifold is a {\em null geodesic}.

The system described by this model is formally integrable \cite{DaHeNi02}.
For every solution $\cV$ giving rise to $\cQ$ and $\cP$ satisfying
(\ref{eom}) we can define a conserved (Noether) charge
\be\label{J}
\cJ = n^{-1} \cV^{-1}\cP\cV,\quad\quad \pt\cJ=0
\ee
taking values in the Lie algebra and obeying a null condition
\be
\langle \cJ | \cJ \rangle = 0.
\ee
Similarly, we can define a (symmetric) `metric' $\cM$ associated with the
$\infty$-bein $\cV$ by $\cM = \cV^T\cV$, which is related to $\cJ$ via
\be
\cJ= \frac12 n^{-1} \cM^{-1}\pt\cM  
\ee
Eqn.~(\ref{J}) implies the following constraint on $\cJ$, and
hence on the initial data, 
\be
\cJ^T=\cM\cJ\cM^{-1} 
\ee
The general solution for $\cM$ can be (formally) written as
\be
\label{genmsol}
\cM(t) = \exp\big(\nu(t) \cJ^T\big) \cM(0) \exp\big(\nu(t)\cJ\big).
\ee
where
\be\label{nu}
\nu(t) := \int_{t_0}^t n(t') dt',
\ee
with $t_0$ some arbitrary initial time.
Translated back to $\cV(t)$, the general solution is
\be\label{cV}
\cV(t) = k(t) \cV(0) \exp\big(\nu(t)\cJ\big).
\ee
Here, $k(t)$ belongs to the compact subgroup $K(AE_n)$ and is not
determined by the equations of motion. This indeterminacy corresponds
to the freedom of choosing a gauge for $\cV(t)$. We can now work out
the above equations of motion (\ref{eom}) level by level, following
the low level results of \cite{DaHeNi03}.

\begin{subsection}{Restriction to $A_{n-2}^{(1)}$}

We will now restrict the one-dimensional geodesic $\s$-model (\ref{Lag}),
to its {\em affine subsector}; this is a consistent truncation.
As we explained, the affine sector originates from the zeroth level
in the $A_{n-2}^{(1)}$ analysis. To this end we define
\be\label{SJ1}
S_{m}^{\a\b} &=& \frac12(\Kb_{m}^\a{}_\b + \Kb_{-m}^\b{}_\a),\nn\\
J_{m}^{\a\b} &=& \frac12(\Kb_{m}^\a{}_\b - \Kb_{-m}^\b{}_\a),\nn
\ee
which obey the symmetry relations
\be\label{SJ2}
S_{m}^{\a\b} = S_{-m}^{\b\a},\quad\quad J_{m}^{\a\b}= - J_{-m}^{\b\a}.
\ee
The generators $S_m^{\a\b}$, $J_m^{\a\b}$ for $m\ge 0$, together
with $\ch$ and $\dh$, define a basis of $A_{n-2}^{(1)}$.
The commutation relations of $A_{n-2}^{(1)}$ in this basis are
\be
[\dh , S_m^{\a\b} ] = - m J_m^{\a\b} \quad , \qquad
[\dh , J_m^{\a\b} ] = - m S_m^{\a\b}
\ee
and, for $m,n\ge 0$,
\be
\left[J_{m}^{\a\b},S_{n}^{\g\d}\right] &=&
\frac12\d^{\b\g}S_{m+n}^{\a\d}
+\frac12\d^{\b\d}S_{m-n}^{\a\g}
-\frac12\d^{\a\g}S_{m-n}^{\d\b}
-\frac12\d^{\a\d}S_{m+n}^{\g\b}\nn\\
&& +\frac12 m \ch \d_{m,n}
\d^{\a\g}\d^{\b\g},\\
\label{jjcom}
\left[J_{m}^{\a\b},J_{n}^{\g\d}\right] &=&
\frac12\d^{\b\g}J_{m+n}^{\a\d}
-\frac12\d^{\b\d}J_{m-n}^{\a\g}
+\frac12\d^{\a\g}J_{m-n}^{\d\b}
-\frac12\d^{\a\d}J_{m+n}^{\g\b},\\
\left[S_{m}^{\a\b},S_{n}^{\g\d}\right] &=&
\frac12\d^{\b\g}J_{m+n}^{\a\d}
+\frac12\d^{\b\d}J_{m-n}^{\a\g}
-\frac12\d^{\a\g}J_{m-n}^{\d\b}
-\frac12\d^{\a\d}J_{m+n}^{\g\b}.
\ee
Observe that the central charge drops out in the $[J,J]$ and $[S,S]$
commutators, but is present in the `mixed' commutators $[J,S]$. The
elements $J_{m}$ (with $m\geq 0$) generate the (centerless)
`maximal compact' subalgebra $K(A^{(1)}_{n-2}) \equiv
K(\widehat{\mf{sl}_{n-1}})$.\footnote{That the latter is {\em not} a
  Kac--Moody  algebra is demonstrated in appendix \ref{compapp}. Note
  also that the additional CSA elements $\ch$ and $\dh$ do not survive
  the projection to the compact subalgebra and hence $K(A_{n-2}^{(1)})$
  equals the compact subalgebra of the loop algebra
  $\widehat{\mf{sl}_{n-1}}$.}

Next, we expand the Cartan form in this basis as in (\ref{QP}), with
\be\label{CF0}
\cQ (t) &=&  Q^{(0)}_{\a\b} J_{0}^{\a\b} + \sum_{m\ge 1} Q^{(m)}_{\a\b}
J_{m}^{\a\b},\nonumber\\
\cP(t) &=&  P^{(0)}_{\a\b} S_{0}^{\a\b} + \sum_{m\ge 1} P^{(m)}_{\a\b}
S_{m}^{\a\b} + \pt\hat\r\, \hat\r^{-1}\, \dh + \pt\s\, \ch.
\ee
The quantity $\rh$ will be seen to be directly related to the `dilaton'
field (=volume density for the internal manifold), while $e^\s$
is related to the conformal factor in the dimensional reduction of
Einstein's theory to two dimensions. Observe that the position of
indices no longer matters, as the tensors appearing on the r.h.s. are
to be regarded as $SO(n-1)$ tensors only. Also,
\be
Q^{(0)}_{\a\b} = - Q^{(0)}_{\b\a} \quad , \qquad
P^{(0)}_{\a\b} = + P^{(0)}_{\b\a}
\ee
whereas the higher modes are not subject to any such symmetry restrictions.

In this general, non-gauge-fixed form, the equations of motion
(\ref{eom}) read
\be
\label{eoms}
n \pt \big(n^{-1}\rh^{-1}\pt \rh\big) &=& 0,\nonumber\\
n \pt\big(n^{-1}\pt\s\big) &=& \frac12 \sum_{m=1}^\infty m
  Q^{(m)}_{\a\b}P^{(m)}_{\a\b} \quad ,\nonumber\\
n \pt\big(n^{-1}P^{(0)}_{\a\b}\big) &=& 2
  Q^{(0)}_{(\a|\g} P^{(0)}_{\g|\b)}
  +\frac12\sum_{m=1}^\infty\big(Q^{(m)}_{(\a|\g|} P^{(m)}_{\b)\g}
  -Q^{(m)}_{\g(\b}P^{(m)}_{|\g|\a)}\big)
\ee
for the level-zero degrees of freedom, where the vertical bars on
the r.h.s. of the third equation indicate that symmetrization should
be performed only over the indices $\a$ and $\b$. At levels $\ell\geq 1$,
we obtain (recall that there is no symmetry under the exchange
of $\a$ and $\b$)
\be\label{eomsl}
n \pt\big(n^{-1} P^{(\ell)}_{\a\b}\big) &=&
\ell \rh^{-1} \pt\rh\, Q^{(\ell)}_{\a\b} +  \frac12 \sum_{m=0}^\ell
\Big( Q^{(m)}_{\a\g} P^{(\ell -m)}_{\g\b}
    -  Q^{(m)}_{\g\b} P^{(\ell - m)}_{\a\g}\Big) \nonumber \\
   && \!\!\!\!\!\!\!\!\!\!\!\!\!\!\!\!\!\!\!\!\!\!\!
\!\!\!\!\!\!\!\!\!\!\!\!\!\!\!\!\!\!\!\!\!
    + \frac12 \sum_{m=0}^\infty \Big( Q^{(\ell +m)}_{\a\g} P^{(m)}_{\b\g}
    - Q^{(\ell + m)}_{\g\b} P^{(m)}_{\g\a}
    + Q^{(m)}_{\b\g} P^{(\ell +m)}_{\a\g}
    - Q^{(m)}_{\g\a} P^{(\ell +m)}_{\g\b}\Big).
\ee
Finally, the Hamiltonian constraint in a general gauge is
\be\label{PP}
\langle\cP | \cP \rangle = - 2 \,\rh^{-1}\pt\rh \, \pt\s +
    P^{(0)}_{\a\b} P^{(0)}_{\a\b} +\frac12
\sum_{m\ge 1} P^{(m)}_{\a\b} P^{(m)}_{\a\b}  = 0.
\ee
The first equation in (\ref{eoms}) can be integrated straightforwardly,
with the result
\be
n(t) = \pt\ln\rh(t) \quad \Rightarrow \quad \nu(t) = \ln\rh(t)
\ee
where integration constants have been chosen conveniently. Hence, the 
choice of the function $\rh(t)$ can be viewed as a choice of gauge
for the lapse $n(t)$; this function is only subject to the requirement 
that it be a monotonic function of the affine parameter $t$. This result 
can be plugged into the general solution (\ref{cV}) to give
\be
\cV(t) = k(t) \cV(0) \exp\big(\ln\rh(t) \cJ\big).
\ee

Although the above equations of motion constitute a consistent
truncation of the full $AE_n/K(AE_n)$ $\s$-model, one may ask how
they embed into the latter. Within $AE_n/K(AE_n)$, the above equations
of motion may receive new contributions from those `non-gradient
fields' whose associated Lie algebra elements `commute back' into the
affine subalgebra. The role and significance of these extra contributions
is not clear.

We emphasize that the expression (\ref{CF0}) is the most general, because
we have not chosen a gauge for the local subgroup  $K(A^{(1)}_{n-2})$.
In other words, {\em no matter which gauge is chosen, the equations of
motion can always be cast into the form (\ref{eoms}), (\ref{eomsl})
and (\ref{PP}).}

\end{subsection}

\begin{subsection}{Triangular gauge}

The equations of motion derived above can now be considered in various
gauges, thereby fixing the factor $k(t)$ in (\ref{cV}). A convenient
choice, and one that has been used almost exclusively in previous
work, is the {\em triangular gauge}, for which the $\infty$-bein $\cV(t)$
depends only on the level $\ell\geq 0$ fields. In this gauge, one obtains
\be\label{CF}
\pt\cV\cV^{-1} = Q^{(0)}_{\a\b} J_{0}^{\a\b} + P^{(0)}_{\a\b}
S_{0}^{\a\b} + \sum_{m\ge 1} P^{(m)}_{\a\b} \Kb_{m}^\a{}_\b +
  \pt\hat\r\, \hat\r^{-1} \, \dh+\pt\s \,\ch.
\ee
Consequently,
\be\label{Q=P}
Q^{(m)}_{\a\b} = P^{(m)}_{\a\b} \qquad \mbox{for $m\geq 1$}.
\ee
Note that the second term on the r.h.s. of (\ref{eomsl}) vanishes for
this choice of gauge, in agreement with previous calculations \cite{DaNi04}.
While $P^{(0)}_{\a\b}$ is symmetric, $P^{(m)}_{\a\b}$ contains both
symmetric and antisymmetric parts for $m>0$, and we can therefore
decompose it into irreducible $\mf{so}_{n-1}$ representations:
\be
P^{(m)}_{\a\b} = \Pb^{(m)}_{\a\b} + \Qb^{(m)}_{\a\b}
\ee
with
\be\label{Borel1}
\Pb^{(m)}_{\a\b} = \Pb^{(m)}_{\b\a},\quad\quad
\Qb^{(m)}_{\a\b} = - \Qb^{(m)}_{\b\a},
\ee
remembering that $\Pb^{(m)}_{\a\b}$ is traceless. With these definitions
the dynamical equations (\ref{eoms}) and (\ref{eomsl}) become
\be
\label{cosvol}
n \pt \big(n^{-1}\rh^{-1}\pt \rh\big) &=& 0,\\
\label{cosconf}
n \pt\big(n^{-1}\pt\s\big) &=& \frac12 \sum_{m\ge 1}
  m \big(\Qb^{(m)}_{\a\b}\Qb^{(m)}_{\a\b} +
   \Pb^{(m)}_{\a\b} \Pb^{(m)}_{\a\b}\big),\\
n \pt\big(n^{-1}\Pb^{(0)}_{\a\b}\big) &=& 2\sum_{m\ge 0}
  \Qb^{(m)}_{(\a|\g}\Pb^{(m)}_{\g|\b)}\; ,\\
n \rh^{\ell}\pt\big(n^{-1}\rh^{-\ell} \Pb^{(\ell)}_{\a\b}\big) &=&
  2\sum_{m\ge 0} \big(\Qb^{(\ell+m)}_{(\a|\g}\Pb^{(m)}_{\g|\b)} +
  \Qb^{(m)}_{(\a|\g}\Pb^{(\ell+m)}_{\g|\b)}\big)\; ,\\
\label{cosgauge}
n\rh^{\ell}\pt\big(n^{-1}\rh^{-\ell} \Qb^{(\ell)}_{\a\b}\big) &=& 2\sum_{m\ge 0}
  \Big(- \Qb^{(\ell +m)}_{\g[\a}\Qb^{(m)}_{\b]\g} +
  \Pb^{(\ell +m)}_{\g[\a}\Pb^{(m)}_{\b]\g}\Big),
\ee
where the last two equations hold for $\ell\geq 1$.
The Hamiltonian constraint (\ref{PP}) now reads
\be
\langle\cP | \cP \rangle = - 2 \,\rh^{-1}\pt\rh \, \pt\s +
    \Pb^{(0)}_{\a\b} \Pb^{(0)}_{\a\b} +\frac12
\sum_{m\ge 1} \big(\Qb^{(m)}_{\a\b}\Qb^{(m)}_{\a\b} + \Pb^{(m)}_{\a\b}
  \Pb^{(m)}_{\a\b}\big) = 0.
\ee

Are there other viable gauge choices? While the triangular or Borel
parametrisation is by no means the only possibility for finite-dimensional
matrices, the situation is more subtle for infinite dimensional groups.
In fact, it appears that so far the triangular parametrization is the
only manageable one for indefinite KMAs. Nevertheless, indications
were found recently \cite{DaNi05} that the triangular parametrization
must be relaxed if one is to include M theoretic corrections, and to
extend the `dictionary' beyond the very first few levels. Here we
simply emphasize again, that independently of the gauge, the equations
of motion can always be written in the form (\ref{eoms}), (\ref{eomsl})
and (\ref{PP}). The only difference is that the equality (\ref{Q=P})
will fail to hold in a non-triangular gauge.

\end{subsection}

\end{section}

\begin{section}{Comparison with two-dimensional reduction}
\label{2dsec}

\begin{subsection}{Relation to higher dimensional vielbein}
\label{vielbeinsec}

The infinite `matrix' $\cV(t) \in AE_n/K(AE_n)$ can be thought of as
an $\infty$-bein analogous to the vielbein of general relativity.
Indeed, it contains the (spatial) vielbein as a finite-dimensional
`submatrix', if one restricts attention to the level zero sector,
{\it i.e.} upon restriction of the full coset to the finite-dimensional
coset space $GL(n)/SO(n)$ (note that we keep the $GL(1)$ factor here as
a remnant of the full algebra). As is well known, pure gravity in
$(n+1)$ spacetime dimensions exhibits a hidden $SL(n-1)$ symmetry
after reduction to three dimensions. This symmetry is obtained through
an enlargement of the manifest $SL(n-2)$ symmetry upon dualization
of $(n-2)$ Kaluza Klein vectors to $(n-2)$ scalar fields. After further
reduction to two dimensions the system becomes integrable with a Lax pair,
and admits an $A_{n-2}^{(1)}$ symmetry as a `solution generating
group' \cite{Julia, BrMa87,Nicolai:1991tt}; for $n=3$ this group is known
as the Geroch group \cite{Geroch,Hoens}. Together with the $GL(n)$ acting
on the spatial $n$-bein, we thus recover precisely the subgroups of $AE_n$
discussed in the foregoing section.

Let us therefore spell out this correspondence in a little more detail,
in order to facilitate the comparison between the one-dimensional
geodesic $\s$-model introduced in the foregoing section, and the
$D=2$ theory to be discussed in the following subsection. For this purpose
we need to fix gauges such that
\be
\label{2dvierbein}
E_M{}^A=
\left(\begin{array}{c|c}N&0\\\hline0&e_m{}^a\end{array}\right).
\ee
This equation displays the decomposition of the $(n+1)$-bein into
a lapse factor $N$, and a spatial $n$-bein $e_m{}^a$; the shift variables
(which enforce the spatial diffeomorphism constraint) have been set to
zero, as required for the comparison with the Kac--Moody $\s$-model.
As we said already, the spatial $n$-bein can be viewed as the level-0
sector of the $\infty$-bein $\cV$, and thereby of the coset $AE_3/K(AE_3)$.
As is well known \cite{DaHeNi03}, the restriction of the of the $\s$-model
Lagrangian (\ref{Lag}) to the $\ell =0$ sector coincides precisely with
the dimensional reduction of Einstein's theory to one time dimension.

In the following section \ref{compare},
we will compare the one-dimensional $\s$-model
equations of motion with those obtained in the reduction of Einstein's
theory to (1+1) dimension (which depend on time and 
one extra spatial coordinate).
For this comparison, one further step is required, namely the split of the
spatial $n$-bein in (\ref{2dvierbein}) according to
\be
\label{2dvierbein1}
E_M{}^A =
\left(\begin{array}{c|cc}\l&0&0\\ \hline 0&\l&0\\
0&0&\r^{1/(n-1)} \bar{e}_\mb{}^\ab \end{array}\right).
\ee
Here we have singled out one (the first) spatial direction, and
decomposed the remaining $(n-1)$-bein into a unimodular part, and
its determinant. So $\det (\bar{e}_\mb{}^\ab)=1$, and $\r$ measures
the volume of the $(n-1)$-dimensional internal space (`internal' from
the (1+1)-dimensional perspective, of course). In addition,
we have adopted the conformal gauge for the zweibein, with the
conformal factor $\l$, thus tying the lapse to the (11) component
of the spatial metric. Finally, setting the remaining elements of the
first column and the first row of $e_m{}^a$ to zero is related to the
hypersurface orthogonality of the Killing vectors usually assumed
in the reduction to (1+1) dimensions \cite{Hoens,BrMa87}.

The special parametrisation (\ref{2dvierbein1})
results in the following non-vanishing components of the coefficients
of anholonomicity $\O_{AB\,C} \equiv 2E_A{}^M E_B{}^N \p_{[M} E_{N]C}$
(with $\partial_x\equiv\partial/\partial{x^1}$)
\be
\O_{0\bb\,\cb} &=& \frac1{n-1}\l^{-1}\d_{\bb\cb} \pt\r \r^{-1}
  + \l^{-1}\eb_\bb{}^\mb\pt \eb_{\mb\cb},\\
\O_{1\bb\,\cb} &=& \frac1{n-1}\l^{-1}\d_{\bb\cb} \p_x\r \r^{-1}
  + \l^{-1} \eb_\bb{}^\mb\p_x \eb_{\mb\cb},\\
\O_{01\,1} &=& \l^{-2} \pt\l,\\
\O_{01\,0} &=& -\l^{-2} \p_x \l.
\ee
The spatial components $\O_{ab\,c}$ of the anholonomicity
are related by duality to the first gradient representation
(\ref{dualgravstart}), which we called dual graviton in
\ref{horgrad}. The irreducibility constraint (\ref{dualgravend})
on this representation is equivalent to the tracelessness
condition $\O_{ab\,b}=0$ \cite{DaHeNi02,DaNi04}. In the present
two-dimensional context this condition reduces to
\be
\label{otr}
\O_{1b\,b} &=& \O_{1\bb\,\bb} = \l^{-1} \p_x \r \r^{-1} = 0.
\ee
Therefore the irreducibility condition $\Omega_{ab\,b} =0$ holds
if and only if $\r=\r(t)$. In the next section, we will arrive at
the same conclusion by a somewhat different route.

The parametrisation (\ref{2dvierbein1}) also yields the determinant of
the original $(n+1)$-bein to be
\be
\det(E)=N\sqrt{g} = \l^{2}\r
\ee
where $\sqrt{g}\equiv \det e_m{}^a$. Using the identification between the
original lapse $N$ and the $\s$-model lapse $n$ derived in
\cite{KlNi04a,DaNi04}, we can relate the $\s$-model lapse to the
internal volume density $\rho$ via
\be\label{n}
n=N g^{-1/2}=\r^{-1}.
\ee
This result will be useful below.

To deduce the relation between the field $\s$ and the conformal
factor $\l$, we compute the contribution from (\ref{2dvierbein1}) to the
diagonal part of $\pt e_a{}^m e_m{}^b$, viz.
\be
\pt e_a{}^m e_m{}^b =
-\l^{-1}\pt \l \, K^1{}_1-\frac1{n-1}\r^{-1}\pt \r \, K^\a{}_\a + \dots
\ee
where the dots stand for the contributions from the off-diagonal degrees
of freedom. Rewriting this in terms of the CSA elements $\ch$ and $\dh$,
cf. (\ref{canddiden}), we obtain
\be\label{cfcart}
\pt[\r^{(n-2)/2(n-1)}\l]\,[\r^{(n-2)/2(n-1)} \l]^{-1}\,\ch
+\pt\r \r^{-1}\,\dh.
\ee
Comparing (\ref{cfcart}) and (\ref{CF0}) we conclude that the conformal
factor should be identified as\footnote{The
  same correction to the conformal factor is also obtained (for $n=3$)
  in the Matzner--Misner coset formulation of gravity reduced to
  two space-time dimensions \cite{BrMa87}, see also \cite{Nicolai2001}.}
\be
\label{confiden}
e^\s \equiv \l \rho^{\frac{n-2}{2(n-1)}}
\ee
while $\hat\r$  indeed agrees the internal volume density $\r$. This
is consistent with the result from the irreducibility constraint
(\ref{otr}) above that $\r$ is a function of time only.

\end{subsection}

\begin{subsection}{$\s$-model equations of motion in (1+1) dimensions}
\label{compare}

Next we write out the equations of motion for the $D=2$ theory
as obtained from an $SL(n-1)/SO(n-1)$ $\s$-model coupled to
gravity in two dimensions, and then compare them to the equations
of motion of the one-dimensional $\s$-model derived in the foregoing
section. This will not only allow us to extend previous results on
the matching, but, more importantly, to exhibit those terms which
do {\em not} match, and where the dictionary needs to be modified
if one is to include higher order spatial gradients. The main advantage
of the (1+1) theory is that the mismatch assumes the simplest possible
form, and can therefore be scrutinized in full detail.

The basic object of the (1+1)-dimensional theory is a matrix $\mt{V} (t,x)$,
which is an element of the coset space $SL(n-1)/SO(n-1)$, and which is
the analog of $\cV(t)$ of the previous section. The corresponding sector
of the theory is governed by the standard $\s$-model Lagrangian on the
worldsheet in a gravitational background. The corresponding Cartan form
belongs to the horizontal $\mf{sl}_{n-1}$
\be\label{Cartan}
\p_\m \mt{V} \mt{V}^{-1}  = \mt{Q}_\m + \mt{P}_\m \equiv
\mt{Q}_{\m\,\a\b} J_0^{\a\b} + \mt{P}_{\m\, \a\b} S_0^{\a\b}
\ee
with the gauge field $\mt{Q}_\m\in \mf{so}_{n-1}$ and
$\mt{P}_\m\in\mf{sl}_{n-1}\ominus\mf{so}_{n-1}$, and
$\mu, \n, \dots = t,x$. The equations of motion read
\be
\label{graveom3}
D_\m(\r \sqrt{-\g} \g^{\m\n}\mt{P}_\n) = 0,
\ee
where $D_\m \mt{P}_\n \equiv \partial_\m \mt{P}_\n - [\mt{Q}_\m , \mt{P}_\n ]$
is the $\mf{so}_{n-1}$ covariant derivative, and $\g_{\m\n}$ is the
worldsheet metric, with inverse $\g^{\m\n}$ and determinant $\g$.
The additional dependence on the `dilaton' $\r$ in this equation is
a remnant of the reduction from higher dimensions. In the conformal gauge
$\g_{\mu\nu} = \l^2\eta_{\m\n}$ (cf. (\ref{2dvierbein1})), this
equation simplifies to
\be
D^\m (\r \mt{P}_\m ) = 0
\ee
where indices are now to be raised and lowered with the
Minkowski metric $\eta_{\m\n}$. Writing out these equations
in terms of $t$ and $x$ components, we get
\be
\label{2dcos}
\r^{-1}\pt(\r \mt{P}_t) - \r^{-1} \p_x (\r\mt{P}_x) =
  [\mt{Q}_t,\mt{P}_t] - [\mt{Q}_x,\mt{P}_x].
\ee
In addition, from (\ref{Cartan}), we have the integrability conditions
\be\label{2dgauge}
  \pt \mt{P}_x -\p_x \mt{P}_t =
   [\mt{Q}_t,\mt{P}_x] - [\mt{Q}_x,\mt{P}_t] \quad , \quad
 \pt \mt{Q}_x-\p_x \mt{Q}_t =[\mt{Q}_t,\mt{Q}_x]+[\mt{P}_t,\mt{P}_x].
\ee
with the gauge connection $\mt{Q}_t , \mt{Q}_x \in  \mf{so}_{n-1}$.
Here we have written out explicitly all covariantizations in order to
facilitate the comparison with the equations of motion (\ref{cosvol})
following from the one-dimensional $\s$-model.

In  conformal gauge, the `dilaton' $\r$ obeys a free field equation, viz.
\be\label{rho}
(\pt^2 - \partial_x^2)\r(t,x) = 0.
\ee
This equation is satisfied in particular if $\r$ equals one of the
two-dimensional coordinates or a linear combination thereof (Weyl canonical
coordinates). Since we are interested in cosmological applications, for which
$\r$ is a timelike coordinate, we choose\footnote{For spacelike $\r$, we
would obtain a variant of the so-called Einstein-Rosen waves, see e.g.
\cite{McCallum}.}
\be\label{r=t}
\r(t,x) = t.
\ee
To see that this choice matches with the one-dimensional $\s$-model, we
substitute (\ref{n}) into the first equation of (\ref{cosvol}), which gives
\be
\r^{-1} \pt^2 \r = 0
\ee
and indeed agrees with (\ref{rho}) if $\r$ is independent of
$x$. Furthermore, independence of $\r$ of $x$ is consistent with the
Young irreducibility constraint in the $\s$-model as explained above
in (\ref{otr}).

The first order equations for the conformal factor $\l$ are just the
Hamiltonian and diffeomorphism constraints, respectively, and read
\be\label{constr1}
\pt\r\, \l^{-1}\pt\l + \partial_x\r\, \l^{-1}\p_x\l &=&
   \frac12 \r \, {\rm Tr} \, \big(\mt{P}_t^2 + \mt{P}_x^2\big), \\
\pt\r\, \l^{-1}\p_x \l + \p_x\r\, \l^{-1}\pt\l &=& \r \, {\rm Tr} \,\mt{P}_t
\mt{P}_x.
\ee
With (\ref{r=t}) they simplify to
\be\label{constr2}
\l^{-1}\pt\l = \frac{t}2 \, {\rm Tr} \,\big(\mt{P}_t^2 + \mt{P}_x^2\big) \quad ,\quad
\l^{-1} \p_x \l = t \, {\rm Tr} \,\mt{P}_t \mt{P}_x.
\ee
There is also a second order equation for the conformal factor,
which reads, for $\r = t$,
\be\label{2dconfsimpl}
\label{2dconfsimpl1}
t^{-1}\pt(t\l^{-1}\pt \l) -\p_x(\l^{-1}\p_x\l) = {\rm Tr}\,\mt{P}_x \mt{P}_x.
\ee

In order to compare the equations (\ref{2dcos})--(\ref{2dconfsimpl})
with the equations of the one-dimensional affine $\s$-model
(\ref{cosvol})--(\ref{cosgauge}), we must truncate both models
appropriately. On the side of the one-dimensional model we do this
by restricting the expansion to levels $|\ell|\leq 1$. Making use of the
equality (\ref{n}) and the triangular gauge (\ref{Borel1}), the equations
of motion (\ref{cosvol})--(\ref{cosgauge}) are thus truncated to
\be
\r^{-1}\pt\Big(\r P^{(0)}_{\a\b}\Big) &=& 2 \Qb^{(0)}_{\g[\a} P^{(0)}_{\b]\g}
    +  2 \Qb^{(1)}_{\g[\a} \Pb^{(1)}_{\b]\g} \nonumber\\
\pt \Pb^{(1)}_{\a\b} &=& 2 \Qb^{(0)}_{\g[\a} \Pb^{(1)}_{\b]\g}
    +  2 \Qb^{(1)}_{\g[\a} \Pb^{(0)}_{\b]\g} \nonumber\\
\pt \Qb^{(1)}_{\a\b} &=& - 2 \Qb^{(1)}_{\g[\a} \Qb^{(0)}_{\b]\g}
    +  2 \Pb^{(1)}_{\g[\a} \Pb^{(0)}_{\b]\g}\\
\r^{-1}\pt (\r \pt\s) &=& \frac12 \left(
   \Qb^{(1)}_{\a\b} \Qb^{(1)}_{\a\b} +
   \Pb^{(1)}_{\a\b} \Pb^{(1)}_{\a\b}\right)
\ee
which must now be matched to (\ref{2dcos}) and (\ref{2dgauge}).
On the side of the (1+1) theory, we must restrict the $x$-dependence
of the two-dimensional quantities such that $\mt{P}_t$ and $\mt{P}_x$
become independent of $x$ (which is tantamount to keeping only first
order spatial gradients). Then the $\s$-model equations of motion
match upon the identification
\begin{align}\label{ID}
\Pb^{(0)}(t) &\equiv \mt{P}_t (t,x_0), &
\Pb^{(1)}(t) &\equiv \mt{P}_x (t,x_0),&\\
\Qb^{(0)}(t) &\equiv \mt{Q}_t (t,x_0),&
\Qb^{(1)}(t)&\equiv - \mt{Q}_x (t,x_0),&
\end{align}
where $x_0$ is some fixed, but arbitrarily chosen spatial point.
The `dictionary' (\ref{ID}) must be supplemented by the relations (\ref{n})
and (\ref{confiden}), already derived before.

While many terms thus do match, there remain several discrepancies between
these equations and those of the geodesic $\s$-model. First of all, there
is a mismatch in the equation of motion for the conformal factor, which
is similar to the one encountered in previous work, and which in particular
involves a contribution $\propto Q^{(1)} Q^{(1)}$, which has no analog
involving the {\em gauge-variant} expression $\mt{Q}_x \mt{Q}_x$ in 
(\ref{constr2}) above. We interpret this mismatch as another indication
of the impossibility to reconcile the higher dimensional gauge
invariance with the desired correspondence. Likewise, (\ref{2dconfsimpl1})
has a term $\p_x^2\s$ which has no analogue in (\ref{cosconf}).
Furthermore, the spatial gradients $\p_x \mt{P}_t, \p_x \mt{P}_x$ and $\p_x
\mt{Q}_t$ in (\ref{2dcos}) are absent in the one-dimensional model.
In view of these mismatches, we will now simplify the affine model
yet further to an exactly solvable model.

\end{subsection}

\end{section}

\begin{section}{Restriction to the `Heisenberg coset' $\H/K(\H)$}
\setcounter{equation}{0}
\label{heissec}

\begin{subsection}{The  $\H/K(\H)$ $\s$-model}
\label{heismod}

We consider the (prototype) affine coset $A_1^{(1)}/K(A_1^{(1)})$
restricted to its `Heisenberg  subspace' $\H/K(\H)$. In this way, 
we are able to define a one-to-one correspondence between a very 
limited, albeit non-trivial, class of solutions of Einstein's 
equations (diagonal metrics with two commuting Killing vectors), 
and the null geodesic motion on the infinite-dimensional manifold  
$\H/K(\H)$. To this end we restrict the affine algebra $A_1^{(1)}$ 
to its Heisenberg subalgebra
\be
{\rm Lie} (\H) := \widehat{\mf{gl}_1} \oplus \reals \ch \oplus \reals \dh
\ee
Note that our terminology is slightly unusual in that the last summand
is usually not considered to be part of the Heisenberg algebra, but
required here in order to obtain a non-degenerate (indefinite) metric
on Lie($\H$). Define
\be
H_m = \frac1{\sqrt{2}}\big(\Kb^2_m{}_2 - \Kb^3_m{}_3\big)
\ee
for all $m\in\ints$. Then the non-vanishing commutators of the
Heisenberg algebra are
\be\label{H1}
\big[H_m, H_n\big] = m \d_{m,-n} \ch,\quad\quad
\big[\dh,H_m\big] = -m H_m.
\ee
The symmetric and antisymmetric combinations are
\be
S_m &:=& \frac12 (H_{m} + H_{-m}) \quad\mbox{for $m\geq 0$}, \nn\\
J_m &:=& \frac12 (H_{m} - H_{-m}) \quad\mbox{for $m\geq 1$}.
\ee
The $J_m$ generate the maximal compact subgroup $K(\H)$ of the extended
Heisenberg group $\H$. The commutation relations (\ref{H1}) in this
basis are (recall that $m,n\ge 0$)
\be\label{H2}
\big[S_m, S_n\big] &=& 0,\quad\quad\quad\quad \big[J_m, J_n\big] =0,
\nonumber\\
\big[J_m, S_n\big] &=& \frac12 m \d_{m,n}\ch,
\nonumber\\
\big[\dh, S_m\big] &=& -m J_m,\quad\quad\!\! \big[\dh, J_m\big] = -m S_m.
\ee
Parametrising the coset space $\H/K(\H)$ in triangular gauge by
\be
\cV (t) =\exp[\ln \rh(t)\, \dh]\cdot \exp[\s(t) \, \ch]\cdot
  \exp\left(\sum_{m\ge 0}\phi_m(t) H_{m}\right),
\ee
we find
\be\label{Vt}
\pt\cV \cV^{-1} = \sum_{m\ge 0} \rh^{-m} \pt\phi_m H_m +\rh^{-1}\pt\rh\,\dh
+\pt\s\,\ch.
\ee
Defining $P_m(t)\equiv  \rh^{-m} \pt\phi_m$, we have
\be
\cP(t) &=& \sum_{m\geq 0} P_m(t) S_m + \rh^{-1}\pt\rh (t)\, \dh +
  \pt\s(t)\,\ch\nn\\
\cQ (t) &=& \sum_{m\geq 1} P_m (t) J_m
\ee
Using the commutation relations (\ref{H2}) we get
\be
[\cQ,\cP] = \sum_{m\geq 0} m P_m \rh^{-1}\pt\rh\, S_m +
\frac12 \sum_{m\geq 1} m P_m^2  \, \ch.
\ee
Therefore the equations of motion of our model read
\be\label{eomHeis}
n\pt(n^{-1}\rh^{-1}\pt\rh) &=& 0,\nonumber\\
n\pt(n^{-1}\pt \s) &=& \frac12 \sum_{m\ge 1}m \rh^{-2m} (\pt\phi_m)^2,
\nonumber\\
n\pt(n^{-1}\rh^{-m}\pt\phi_m) &=& m \rh^{-(m+1)}\pt\rh\,\pt\phi_m.
\ee
The Hamiltonian constraint takes the form
\be\label{hamheis}
-2\rh^{-1}\pt\rh\pt\s + P_0^2 +\frac12 \sum_{m\ge 1} P_m^2 =0.
\ee

Using the insights from the preceding section we solve the first
equation of (\ref{eomHeis}) by setting $n^{-1}=\rh=t$. Then the null
geodesic trajectory in $\H/K(\H)$ is explicitly parametrized by
\be
\label{traj}
&&\left.\begin{array}{rcl}\phi_0(t) &=& p_0 \ln t + q_0\\
\phi_m(t) &=& \frac1{2m}p_m t^{2m} + q_m\;\;\; (m>0)
\end{array}\right\} \quad  \Rightarrow \;\; P_m(t) = p_m t^{m-1} \;\;
(m\geq 0) \; ,\nonumber\\
&&\;\;\;\;\;\s(t)\;\; =\; \frac12 p_0^2\ln t +
\sum_{m \ge 1} \frac1{8m} p_m^2 t^{2m} + \s_0.
\ee
where the coefficient of $\ln t$ in the last line is determined
by imposing the Hamiltonian constraint (\ref{hamheis}), which,
however, does not fix $\s_0$. Because the model is explicitly solvable,
we see in particular how the solution $\cV(t)$ depends on the most
general initial data, {\it i.e.} the initial `coordinates' $q_m$
and `momenta' $p_m$ for $m\geq 0$.

The conserved current is, from (\ref{J}),
\be
\cJ &=& n^{-1} \cV^{-1} \cP\cV =  n^{-1}\bigg[\Big(\pt \s-\frac12
\sum_{m\geq 1}  m \rh^{-m} \phi_m P_m \Big)\,\ch +
\rh^{-1}\pt\rh\;\dh\;\\
&& \quad+  P_0 H_0+ \frac12 \sum_{m\geq 1}
\big( - 2m\rh^{-1}\pt\rh\phi_m + \rh^m P_m \big)
H_{m} + \frac12 \sum_{m\geq 1} \rh^{-m} P_m H_{-m}\bigg].\nonumber
\ee
Plugging the solution (\ref{traj}), as well as $n^{-1}(t)=t$, into this
expression yields
\be
\cJ&=&\left(\frac12p_0^2-\frac12 \sum_{m\ge 1} m p_m q_m\right) \ch +
\;\dh\; \nn\\ &&
+\frac12\sum_{m\geq 1} p_m H_{-m} + \, p_0 H_0 \, -\sum_{m\geq 1} m q_m
H_{m},
\ee
which is evidently conserved since it depends only on the initial
data. Observe that neither $q_0$ nor $\s_0$ appear in $\cJ$. In
agreement with the general analysis of \cite{DaHeNi03}, the initial
momenta appear  in the {\em lower}, and the initial coordinates in the
{\em upper}  triangular half of Lie$(\H)$.

Under global $\H$ transformations $g$, $\cJ$ changes as
$\cJ\rightarrow g \cJ g^{-1}$. Let
\be
g(\a,\b,\o): = \exp(\ln\a\, \ch) \cdot \exp(\ln \b \,\dh)\cdot
\exp\left(\sum_{n\in \ints} \o_n H_n\right),
\ee
then
\be
\label{Jtrm}
g \cJ g^{-1} &=& \left(\frac12 p_0^2 - \frac12\sum_{m\ge 1}
m(p_m-2m\o_{-m})(q_m-\o_m)\right)\,\ch + \dh\\
&& - \sum_{m\ge 1} m\b^{-m}(q_m-\o_m) H_m + p_0 H_0
+\frac12\sum_{m\ge 1}\b^m(p_m-2m\o_{-m})H_{-m}.\nn
\ee
From this we immediately read off the transformation of the initial
coordinates $q_m$ and momenta $p_m$ (for $m\ge 1$) under the action
of $\H$:
\be
\label{trmpq}
\left.\begin{array}{lcl}q_m &\to& \b^{-m}(q_m-\o_m)\\
p_m &\to& \b^{m}(p_m-2m\o_{-m})\end{array}\right\}\quad\quad (m\ge 1).
\ee
On $\cV$ the corresponding transformation is $\cV\to k\cV g^{-1}$ where
$k(t)$ is the (local) compensator required to restore triangular gauge,
which does not contribute to the transformation of $\cJ$. By considering
$\cV$ transformations we find constant shifts (which drop out in
$g\cJ g^{-1}$)
\be\label{qsigma}
\s_0  \to \s_0 - \a \quad, \qquad q_0 \to q_0-\o_0.
\ee
Hence, all the constants are shifted except the (Kasner) coefficient $p_0$. 
All fields except $\s$ are inert under a transformation associated 
with the central term. This is no longer the case for the full $AE_n$ 
model where $\ch$ ceases to be central.

\end{subsection}

\begin{subsection}{Relation to polarised Gowdy cosmologies}

Remarkably, there is a one-to-one correspondence between our
model, and the `polarised' Gowdy type cosmological model with
diagonal metrics depending on two coordinates $(t,x)$ \footnote{See
also \cite{BV} for a discussion of such solutions in the framework
of gravitational solitons.}. Apart from possible reparametrisations 
of the time parameter $t$, {\em this correspondence works only after 
complete elimination of the gauge degrees of freedom on both sides.} 
In particular, the conformal factor must be treated as a {\em dependent} 
degree of freedom via the constraints (\ref{lambda}) below. The relevant
line elements can be written in the form (see e.g. \cite{McCallum})
\be\label{ds2}
ds^2 = \lambda^2 e^{-Z}(-dt^2 + dx^2) + t^2 e^{-Z} dy^2 + e^Z dz^2
\ee
where the function $Z=Z(t,x)$ is subject to the two-dimensional
wave equation
\be\label{Gowdy}
\pt \big[t\pt Z(t,x)\big] = t\partial_x^2 Z(t,x)
\ee
The conformal factor can be determined by straightforward integration
from the first order equations
\be\label{lambda}
\l^{-1}\pt\l = \frac{t}4\big[ (\pt Z)^2 + (\partial_x Z)^2\big] \quad , \qquad
\l^{-1}\p_x \l = \frac{t}2 \pt Z \partial_x Z
\ee
which are compatible if (\ref{Gowdy}) is satisfied. They give rise to the
second order evolution equation for $\l$
\be
\pt(t\l^{-1}\pt\l) -t \p_x (\l^{-1} \p_x\l) = \frac{t}2 (\p_x Z)^2.
\ee
The general solution of (\ref{Gowdy}) can be written as \footnote{For 
{\em regular} initial data, {\it i.e.} $p(x)=0$, this formula can be 
written in the completely explicit (and manifestly causal) form \cite{CH}
$$
Z(t,x) = \frac1{\pi} \int_{x-t}^{x+t}
  \frac{q(w)dw}{\sqrt{t^2-(x-w)^2}}.
$$
which also follows directly from (\ref{Bessel}) (we thank V.~Moncrief 
for a discussion on this point).}
\be\label{Green}
Z(t,x) = \int_{-\infty}^\infty G(t,x-w) \tilde{q}(w) dw  
         + \int_{-\infty}^\infty H(t,x-w) p(w) dw
\ee
with the Green's functions
\be\label{Bessel}
G(t,x) := \frac1{\pi} \int_{0}^\infty \cos(kx) J_0(kt) dk \;\; , \quad
H(t,x) := \frac1{\pi} \int_{0}^\infty \cos(kx) Y_0(kt) dk
\ee
where $J_0(z)$ and $Y_0(z) = J_0(z) \ln (z) + \dots$ are the standard 
Bessel functions (see e.g. \cite{Bessel}). Note that the second integral 
in (\ref{Bessel}) is well defined (as a distribution) for all $t>0$ despite 
the logarithmic singularity of the integrand. Near the singular hypersurface 
$t=0$, the general solution admits the expansion \cite{CIM90,KiRe98} 
\be\label{generalZ}
Z(t,x) = q(x) + p(x) \ln t + F(t,x)
\ee
where $q(x) \neq \tilde{q}(x)$ unless $p(x)=0$, and
$\lim_{t\rightarrow 0} F(t,x) = 0$. %($F(t,x)$ contains at most
%one power of $\ln t$ when expanded in $t^m (\ln t)^n$; 
The conformal factor expands as
\be\label{generalL}
\ln\l(t,x) = \frac14 p^2(x) \ln t + \dots
\ee
The functions% $q(x)$ and $p(x)$ with
\be\label{initial}
q(x) = q_0 + q_1 x + q_2 x^2 + \dots \quad , \quad
p(x) = p_0 + p_1 x + p_2 x^2 + \dots
\ee
can be viewed as the {\em initial values of the coordinates and momenta
on the `big bang' hypersurface $t=0$}. $p(x)$ correspond to the
canonical momenta (rather than the velocities) since we know from
section \ref{compare} that the Lagrange density contains the kinetic
term $\cL = \r \mt{P}_t\mt{P}_t$. With $\r(t)=t$ the conjugate momenta 
are therefore
\be
\Pi(t,x) = t\pt Z(t,x) \quad\stackrel{t\to 0}{\longrightarrow}\quad p(x).
\ee
We emphasize that we are here
working locally in a fixed coordinate chart --- the Taylor expansion
could be equivalently performed about any other spatial point $x_0$.

Evidently, the {\em Kasner solution} corresponds to constant
$q(x) = q_0$ and $p(x) = p_0$ in (\ref{generalZ}) and (\ref{generalL}).
In this case we have a direct correspondence between this particular
solution of Einstein's equations, and the corresponding solution (\ref{traj})
of the $\H/K(\H)$ model with $p_m= q_m= 0$ for $m\ge 1$. In view of
(\ref{trmpq}) and (\ref{qsigma}), different orbits under $\H$ are thus
labelled by $p_0$. If $p(x) \equiv 0$, the solution is no longer of
Kasner type; instead of a singularity, it may exhibit other features
such as Cauchy horizons \cite{CIM90}. In the following we will not
consider this case.

When the higher modes (and the higher order terms in
(\ref{generalZ})) are switched on, the relation is less transparent.
{\em We now define an explicit one-to-one map between the
null geodesic trajectories $\cV(t)$ on $\H/K(\H)$ characterized by
(\ref{traj}), and the solutions of (\ref{Gowdy}) characterized by the
initial values (\ref{initial}), by associating the solutions
with the same initial data $q_0,q_1,\dots$ and $p_0, p_1,\dots$ and
$\s_0$.} We emphasize again that such an association makes sense {\em
  only  after} gauges have been fixed; in particular, all possible
coordinate transformations in $(t,x)$ are `used up' when the
four-dimensional line element is cast into the form (\ref{ds2}), and
the diffeomorphism constraint is imposed on the initial data by
solving (\ref{lambda}). The action of the `solution generating group'
$\H$ on the solutions of (\ref{Gowdy}) is likewise {\em defined} by
the corresponding action on the solution $\cV(t)$ in (\ref{Jtrm}). We
next examine how this action compares to the standard realisation of
the Geroch group on such solutions.

\end{subsection}

\begin{subsection}{Geroch vs. Heisenberg}

To elucidate the relation between the transformations (\ref{trmpq})
on the initial data (\ref{initial}) on the one hand, and the action of
the corresponding subgroup of the Geroch group on solutions of (\ref{Gowdy}) 
on the other, we briefly recall how the latter is usually realized. 
The integrability of general relativity reduced to $(1+1)$ dimensions 
\cite{Maison,BZ} is usually expressed by means of a linear system 
(or Lax pair) whose integrability condition is equivalent to the 
reduced Einstein equations \cite{BrMa87}. In the standard formulation
this is achieved by promoting the matrix $\mt{V}(t,x)$ appearing in
(\ref{Cartan}) of section~\ref{2dsec} to an element of the associated
{\em loop group} by introducing an extra dependence on a spectral 
parameter $\g$, viz.
\be
\mt{V}(t,x) \ra \Vh(t,x;\g),
\ee
This matrix satisfies the linear system (Lax pair) equations
\cite{BrMa87,Nicolai:1991tt}
\be
\label{LScart}
\p_\m \Vh \Vh^{-1} = \mt{Q}_\m +\frac{1+\g^2}{1-\g^2} \mt{P}_\m
+\frac{2\g}{1-\g^2} \e_{\m\n}\mt{P}^\n.
\ee
The parameter $\g$ is to be interpreted as the spectral parameter of
the loop algebra over $\mf{sl}_{n-2}$. The integrability of (\ref{LScart})
implies the equations of motion (\ref{2dcos}) if
\be
\label{spec}
\g=\g(t,x;w) = \frac{(x-w) \pm \sqrt{(x-w)^2 - t^2}}{t} \;\;\Leftrightarrow\;\;
w = -\frac{t}2 \left( \g + \frac1{\g}\right) + x
\ee
where $w$ is sometimes called the `constant spectral parameter'. The
appearance of {\em two} spectral parameters $\g$ and $w$, one of which
depends on the space-time coordinates, is a consequence of the coupling the
$\s$-model to gravity, and the characteristic feature which distinguishes
this model from flat space $\s$-models. It is precisely the coordinate
dependence of the spectral parameter $\g$ which allows the Geroch group
to generate space-dependent solutions, as we will now explain for the
polarised Gowdy cosmologies.

Diagonal solutions can be obtained by starting from the following
$w$-dependent matrix \footnote{The following considerations are based
on unpublished work with T.~Damour \cite{DN}.} considered as an element
of the loop group $\widehat{GL(1)}\subset A_1^{(1)}$
\be
\Vh(w)=\left(\begin{array}{cc}\exp(G(w))&0\\0&\exp(-G(w))\end{array}\right)
\ee
To generate space-time dependent solutions from this matrix, one follows
the general procedure of \cite{BrMa87}, by first expressing $w$ as a function
of $t,x$ and $\g$ by virtue of (\ref{spec}), and then removing the
singularity at $\g=0$ by a compensating transformation. The corresponding
solution of the field equations is then obtained from $\Vh$ by setting
$\g=0$, as $\Vh$ is holomorphic at $\g=0$ after removal of the poles. In
infinitesimal form, these steps are summarized in the combined transformation
\be\label{dVh}
\delta\Vh (t,x;\g) = \d h(t,x;\g) \Vh (t,x;\g) - \Vh (t,x;\g) \d g(w)
\ee
where $\d g(w)\in\widehat{\mf{gl}_1}$, and $\d h$ is the compensating
transformation in $K(\widehat{\mf{gl}_1})$ which removes the poles at
$\g=0$. (\ref{dVh}) shows that {\em only half of} $\widehat{\mf{gl}_1}$
has an actual effect on the solution $\mt{V}(t,x)$, namely those
$\d g(w)\propto w^n$ with $n>0$. The other half ($n<0$) yields expressions
in $\d\Vh$ for which $\d\mt{V} = \lim_{\g\rightarrow 0} \d\Vh(\g)=0$.
Hence, the latter transformations merely shift the integration constants
arising in the definition of the higher order dual potentials, and have
no effect on the physical solution.

All solutions of the type (\ref{generalZ}) can be generated from \cite{DN}
\be
\label{generalG}
G(w) = f(w) + g(w) \ln w
\ee
with regular functions $f(w)$ and $g(w)$ (which are directly related to,
but not identical with, the functions $q(w)$ and $p(w)$ in (\ref{initial})).
The presence of a $\ln w$ term in (\ref{generalG}), which is necessary
to obtain a non-trivial Kasner coefficient $p(x)$ in (\ref{generalZ}),
signals that we are not dealing with the standard loop group in $w$.
The necessity of the $\ln w$ term is another manifestation of the fact
that the standard realisation of the Geroch group affects only one half
of the initial data, and must therefore be `amended'.

From (\ref{spec}) and (\ref{generalG}), we can deduce the leading
contributions to $\d Z(t,x)$ for $t\sim 0$ from a given $\d g(w)$, 
following the steps described above, with the result
\begin{align}\label{dZ}
\d g(w)&=w^n&\Rightarrow\quad \d Z(t,x)&\propto x^n,&\nonumber\\
\d g(w)&=w^n\ln w&\Rightarrow\quad \d Z(t,x)&\propto x^n \ln t,&
\end{align}
for $n\ge 0$. This shows that a regular (non-logarithmic) affine level
$n$ element of the Geroch group induces a spatial dependence $\propto x^n$
in the regular initial data in a Taylor expansion around $x=0$, but that
a logarithmic dependence on the loop parameter $w$ is required to change
the singular initial data (the Kasner coefficient function $p(x)$).
The action of the Witt--Virasoro algebra, and hence that of infinitesimal
variations along $\dh$, were discussed in \cite{Julia:1996nu}. The central
charge acts on the conformal factor $\l$ by scaling transformations
\cite{Julia,BrMa87,Nicolai:1991tt}, in agreement with the results 
derived at the end of section~\ref{heismod}.

Eqs.~(\ref{dZ}) illustrate the main advantage of the affine coset model
and our new realization of (a subgroup of) the Geroch group in 
comparison with the linear system approach: In the Heisenberg model,
the {\em full} algebra  generates non-trivial transformations on the
initial data --- the positive half shifts the initial coordinates
$q_m$, and the negative half shifts the initial momenta $p_m$,
see (\ref{trmpq}). By contrast, for the standard realisation of
the Geroch group only {\em half} the transformations $\propto w^n$ 
for $n>0$, act non-trivially and change the initial coordinates encoded 
in $q(x)$; in this sector, the action of the Geroch group agrees with 
the action of the upper half of $\H$. On the other hand, in the standard
approach, we must extend the Geroch transformations by allowing $w^n \ln w$
terms in order to shift the initial momenta $p(x)$. Moreover, this 
rather {\it ad hoc} extension of the loop group fails at the non-linear 
level, {\it i.e.} for non-diagonal metrics because it would force us
to admit {\em arbitrary} positive powers of $\ln w$ \cite{DN}. This difficulty 
is avoided altogether in our new coset approach, where the `solution 
generating group' acts on {\em all} initial data, except $p_0$, and which 
furthermore does not require modifying the loop group by logarithimic
terms to generate the most general solution.

The difference between the two realisations of the affine symmetry
is also evident from way the (dynamical) fields are encoded in $\Vh$ and
$\cV$ respectively. Since the Chevalley involution acts by \cite{BrMa87}
\be
\omega \big(\Vh(t,x;\g)\big) =
\left(\Vh^T\right)^{-1} \left(t,x;\frac1{\gamma}\right)
\ee
the r.h.s. of (\ref{LScart}) is invariant, whence
\be
\p_\m \Vh \Vh^{-1} \in K(A_{n-2}^{(1)}).
\ee
By contrast, for the geodesic $\s$-model of section
\ref{sigmamod}, all the dynamics is contained in the coset components,
see (\ref{CF0}). Hence, we conclude again that the linear system (\ref{LScart})
is not immediately suitable for the comparison with the results of
the foregoing sections.

\end{subsection}

\end{section}

\begin{section}{Discussion}
\setcounter{equation}{0}
\label{concl}

There is an alternative formulation of the Lax pair for two-dimensional
gravity \cite{BeJu99}, which is somewhat more similar to
(\ref{CF}), in that its r.h.s. belongs to a coset subalgebra,
and involves the CSA generators $\ch$ and $\dh$ explicitly. In that
formulation the Cartan form reads (using light-cone coordinates)
\be
\label{BJLS}
\p_\pm\cV \cV^{-1} (t,x) = \pm \r^{-1}\partial_\pm\r (L_0-L_{\pm 1}) +
\frac12 Q_{\pm\a\b} J^{\a\b} +\frac12 P_{\pm\a\b} K_{\pm 1}^{\a}{}_\b \mp
\p_\pm\s\ch.
\ee
Here, $L_0$ and $L_{\pm 1}$ belong to the M\"obius subalgebra of
the Virasoro algebra (which can be embedded in the enveloping
algebra of the affine algebra with $L_0=\dh$). In
contradistinction to the linear system (\ref{LScart}) the
expansion of (\ref{BJLS}) is truncated at affine level one,
but involves the M\"obius generators. (This has an effect on the
set of allowed dressing tranformations \cite{BeJu99}.)

In the above linear system, the generator $L_{-1}$ acts like a derivative
operator on a vertex representation, and belongs to a Witt--Virasoro
algebra acting on the affine algebra via a semi-direct product.
This suggests that there might exist a similar, but larger, subalgebra
inside the enveloping algebra of $AE_n$, which would contain derivative
operators in all spatial directions, and contain the standard Witt--Virasoro
algebra as a subalgebra. In addition, the affine algebra might be
embedded in a {\em toric algebra} (see e.g. \cite{Rao} and references
therein). Such an algebra is spanned by generators $T_{\vec{m}}^A$ (where
$\vec{m}$ designates a vector in some vector space), with commutation
relations of the form
\be
\big[T_{\vec{m}}^A, T_{\vec{n}}^B\big] = f^{AB}{}_C
T_{\vec{m}+\vec{n}}^C + \mbox{central terms}.
\ee
However, looking at the relation for the gradient fields derived for $AE_3$
in section \ref{combaffin}, we see that the naive identification of
$\vec{m}$ with the three-vector of the multiplicities of the indices
$1,2,3$ among the `gradient indices' $a_1,\ldots a_{\ell-1}$ of the
generator $E_{a_1,\ldots,a_{\ell-1}}{}^{b_1b_2}$ will fail since the
commutator (\ref{eecomm}) will always {\em add} one index and so destroy the
vector space structure on such $\vec{m}$. Moreover we have seen in
section~\ref{combaffin} that the gradient representations by themselves
do not close into a subalgebra of $AE_3$. We thus conclude that there
is no toric algebra inside $AE_n$ or its enveloping algebra.

The interpretation of the affine coset proposed in section
\ref{heissec} provides a more favourable realisation of the Geroch
group since it can incorporate transformations of both the initial
coordinates and the initial velocities in the general case without
extending the set of allowed transformations. We anticipate that
similar results hold for the $\s$-model realisation of the affine
group $A_1^{(1)}$ in comparison with the standard realisation of
the full Geroch group. However, the more important challenge at this
point is now to find out how the hyperbolic $AE_3$ coset model can
generate the most general dependence on four-dimensional space and
time coordinates, and to understand the significance of the fact that,
unlike the affine elements, the gradient generators no longer form 
a closed subalgebra of $AE_n$.

It is also interesting to consider higher levels of the
affine coset model by including as a next step the basic representation
\cite{FeFr83} of the affine algebra. This will show new aspects of the
correspondence characteristic for the KMA $\s$-model crucial for the
programme of \cite{DaHeNi02}.

\end{section}

\vspace*{0.5cm}
\noindent{\bf Acknowledgements:}
We are grateful to T.~Damour, F.~Englert, V.~Moncrief and A.~Rendall
for enlightening discussions, to D.~Bernard and T.~Fischbacher for
correspondence, and to T.~Damour for permission to include some
unpublished results of \cite{DN}.

\newpage

\appendix

\begin{section}{The Compact Subalgebra is not a KMA}
\setcounter{equation}{0}
\label{compapp}

In this appendix we prove that the `maximal compact' invariant subalgebra 
of an infinite-dimensional KMA is {\em not} of Kac--Moody type. This 
applies in particular to the algebras $AE_n$ and $E_{10}$, and their
invariant subalgebras.

\begin{prop} Let $\lag$ be any infinite-dimensional split real Kac--Moody 
  algebra $\lag$, which is not the direct sum of (infinitely many)
  finite dimensional algebras. Then the infinite-dimensional
  subalgebra $\lak\subset\lag$ 
  fixed by the Chevalley involution $\o$ is not a Kac--Moody algebra.
\end{prop}

{\bf Proof:} The main observation is that the contravariant Hermitian
form on the Kac--Moody algebra
\be
(x|y) := -\langle x | \o(y)\rangle\quad\quad{\rm for}\,x,y,\in\lag
\ee
is not positive definite everywhere on the split real KMA
if the latter is infinite-dimensional \cite{Ka90}: in the CSA, and only
in the CSA, there exist elements of both positive and negative norm
squared, because the Cartan matrix is not positive definite. However,
since the CSA is not invariant under $\o$, the compact subalgebra $\lak$
does not contain any such elements, and the contravariant form on $\lak$
(inherited from $\lag$) is therefore positive definite on $\lak$ \cite{KP}. 
Because $\lak$ is infinite-dimensional, it can thus be a KMA if and only 
if it is the (orthogonal) sum of infinitely many finite-dimensional 
algebras of KM type.

We will prove that this cannot happen by exhibiting a contradiction if
the assumption were true. Therefore assume that $\lak$
is the infinite direct sum of some finite-dimensional algebras.
Now consider some (regular)
level decomposition $\lag=\bigoplus_{\ell\in\ints} \lag_\ell$ of
$\lag$ such that  $0<\mbox{dim}\,\lag_\ell<\infty$ for all $\ell\in
\ints$. Then $\lag$ is generated by taking multiple commutators of
$\lag_{-1}$, $\lag_0$ and $\lag_{1}$. Since the Chevalley involution 
$\o$ acts according to $\o(\lag_\ell)=\lag_{-\ell}$, the invariant 
subalgebra admits the decomposition (not a grading!)
\be
\lak=\bigoplus_{\ell\geq 0} \lak_\ell \quad\mbox{with}\quad
\lak_0\subset \lag_0,\quad\quad \lak_\ell\subset \lag_{-\ell}\oplus \lag_\ell
\quad \Rightarrow \quad [\lak_0,\lak_\ell]\subset \lak_\ell
\ee
Like $\lag$, the infinite-dimensional invariant subalgeba $\lak$ is 
generated by taking multiple commutators of $\lak_0$ and $\lak_1$.
Since all $\lag_\ell$ are of finite dimension, it follows in particular
that $\lak_0$ and $\lak_1$ are also finite-dimensional.
According to our assumption they must therefore belong to a
{\em finite} sum of finite-dimensional subalgebras of $\lak$.
Hence, the algebra they generate is also finite-dimensional, in
contradiction with the fact that $\lak$ is infinite-dimensional.
$\square$

The proposition unfortunately does not give an answer to the
important question what kind of Lie algebra $\lak$ is, and, more
pressingly, what its representation theory is. It is expected
that the representation theory of $\lak$ is associated
with the supersymmetric extension of the one-dimensional $\s$-model
\cite{NiSa04,KlNi04a}. 

\end{section}

\begin{section}{$AE_n$ $\s$-model in $\mf{sl}_n$ decomposition and
    $p$-forms}
\setcounter{equation}{0}
\label{pformapp}

In this appendix, we analyse the $\s$-model based on $AE_n$ of section
\ref{sigmamod} from its $\mf{sl}_n$ subalgebra. We will also add
$p$-forms to the Einstein equation of the model to
make contact with other oxidized theories.

\begin{subsection}{Pure Gravity}

We first study the $\s$-model and then compare it to the reduction of
the Einstein equation.

The Cartan form in triangular gauge in $\mf{sl}_n$ form now is
\be
\label{peinstcart}
\pt \cV \cV^{-1} &=& P_{ab} S^{ab} + Q_{ab}J^{ab}+ \frac1{(n-2)!}
P_{a_1\ldots a_{n-2},a_{n-1}} E^{a_1\ldots a_{n-2},a_{n-1}}+\ldots
\ee
with indices $a,b$ taking values $1,\ldots,n$. The resulting
$\s$-model equation on level $\ell=0$, in the truncation to $|\ell|\le
1$ can be deduced from the results of section \ref{slnsub}, with the result
\be
\label{peinsteq}
&&n\pt(n^{-1} P_{ab}) = 2 P_{c(a}Q_{b)c} +
\frac1{2(n-2)!}\bigg(\d_{ab} P_{c_1\ldots c_{n-2},c_{n-1}}P_{c_1\ldots
  c_{n-2},c_{n-1}} \nn\\&&\quad\quad-P_{c_1\ldots c_{n-2},a}P_{c_1\ldots c_{n-2},b}
-(n-2) P_{ac_1\ldots c_{n-3},c_{n-2}}P_{bc_1\ldots c_{n-3},c_{n-2}}\bigg).
\ee

In the corresponding gravity theory, the space-space components of
reduced Einstein equations in flat components are
\be
\label{peinstsys}
R_{ab}= R^{\rm{temp}}_{ab} + R^{\rm{spat}}_{ab} = 0,
\ee
where we defined, following \cite{DaNi04},
\be
R^{\rm{temp}}_{ab} &=& \p_0\o_{ab0}+\o_{cc0}\o_{ab0}-2\o_{0c(a}\o_{b)c0},\\
\label{riccispat}
R^{\rm{spat}}_{ab} &=&
\frac14\tO_{cd\,a}\tO_{cd\,b}-\frac12\tO_{ac\,d}\tO_{bc\,d}
-\frac12\tO_{ac\,d}\tO_{bd\,c} -\frac12\p_c\tO_{c(a\,b)}.
\ee
Here, $\tO_{ab\,c}$ is the tracefree part of the anholonomy
$\Omega_{ab\,c}=2 e_{[a}{}^m e_{b]}{}^n \p_m e_{nc}$ and
$\o_{a\,bc}=(\Omega_{ab\,c}+\Omega_{ca\,b}-\Omega_{bc\,a})/2$ is the
spin connection. We use the same decomposition (\ref{2dvierbein}) of the
vielbein as in section \ref{vielbeinsec}.

By comparing eq.~(\ref{peinsteq}) with eq.~(\ref{peinstsys}), we see
that almost all terms match upon identifying
\be
\label{einstid}
n(t) &\equiv& N e^{-1}(t,x_0),\\
P_{ab} (t) &\equiv& \o_{a\,bt} (t,x_0),\\
Q_{ab} (t) &\equiv& -\o_{t\,ab} (t,x_0),\\
P_{a_1\ldots a_{n-2},a_{n-1}}(t) &\equiv& \frac{N}{2}\ve_{a_1\ldots a_{n-2}
  bc}\tO_{bc\,a_{n-1}}(t,x_0)
\ee
for some fixed spatial point $x_0$. The terms which do not match are
the second spatial derivatives and the cross-term
$\tO_{ac\,d}\tO_{bd\,c}$ in (\ref{riccispat}). However, the second
spatial derivatives are not expected to match in this approximation
and the cross-term disappears upon using a conformal gauge for the
zweibein.

\end{subsection}

\begin{subsection}{Inclusion of $p$-forms}

We now add a single $p$-form field to the Einstein action via their kinetic
terms\footnote{We need not concern ourselves with possible
  Chern--Simons-like terms since these do not affect the Einstein
  equation. However, they are crucial for the gauge field equation of
  motions and Bianchi identities.}
\be
\label{einstact}
S=\int d^{n+1}x \det(E)\bigg( R - \frac1{2\,(p+1)!} F_{M_1\ldots M_{p+1}}
F^{M_1\ldots M_{p+1}}\bigg).
\ee
The generalisation to several $p$-forms is straight-forward and we do
not consider dilaton terms here for simplicity.

The Einstein equation resulting from (\ref{einstact}) can be written as
\be
R_{MN}=\frac1{2\,p!}F_{MP_1\ldots P_p}F_{N}{}^{P_1\ldots P_p}
-\frac{p}{2\,(p+1)!\,(n-1)}G_{MN} F_{P_1\ldots P_{p+1}} F^{P_1\ldots
  P_{p+1}}.
\ee
Converting into flat indices and restricting to the spatial
directions, we write
\be
\label{einsteq1}
R^{\rm{temp}}_{ab} + R^{\rm{spat}}_{ab} = -N^{-2} T^{\rm{el}}_{ab} +
T^{\rm{magn}}_{ab},
\ee
with the energy momentum tensor split into `electric' and `magnetic'
contributions according to
\be
\label{elstress}
T^{\rm{el}}_{ab} &=& \frac{p}{2\,p!}F_{tac_1\ldots c_{p-1}}
F_{tbc_1\ldots c_{p-1}} - \frac{p}{2\,p!\,(n-1)}\d_{ab} F_{tc_1\ldots
  c_p} F_{tc_1\ldots c_p},\\
T^{\rm{magn}}_{ab} &=& \frac{1}{2\,p!}F_{ac_1\ldots c_p}
F_{bc_1\ldots c_p} - \frac{p}{2\,(p+1)!\,(n-1)}\d_{ab} F_{c_1\ldots
  c_{p+1}} F_{c_1\ldots c_{p+1}}.
\ee
The factor $-N^{-2}$ in (\ref{einsteq1}) stems from lowering the $t$
index. Similarly, we split the Ricci tensor into temporal and spatial
derivatives as before.

On the algebra side, we need to include additional generators to
account for the electric and magnetic contributions to the Einstein
equation. Consider the expansion of the Cartan form to be
\be
\label{einstcart}
\pt \cV \cV^{-1} &=& P_{ab} S^{ab} + Q_{ab}J^{ab}+ \frac1{(n-2)!}
P_{a_1\ldots a_{n-2},a_{n-1}} E^{a_1\ldots a_{n-2},a_{n-1}}\\
&&\quad + \frac1{p!} P^{\rm{el}}_{a_1\ldots a_p} E^{a_1\ldots a_p} +
\frac1{(n-p-1)!} P^{\rm{magn}}_{a_1\ldots a_{n-p-1}} E^{a_1\ldots
  a_{n-p-1}} +\ldots.\nn
\ee
In the first line of (\ref{einstcart}), we have the contribution of
the $AE_n$ fields as above. The second line now belongs to additional
generators we have introduced into the algebra. That such generators
are present in the relevant Kac--Moody algebra for the oxidized theory follows
from the results of \cite{KlSchnWe04}. Demanding the normalizations
\be
\langle E^{a_1\ldots a_p} | F_{b_1\ldots b_p} \rangle &=& p!\,
\d^{a_1\ldots a_p}_{b_1\ldots b_p},\\
\langle E^{a_1\ldots a_{n-p-1}} | F_{b_1\ldots b_{n-p-1}} \rangle &=&
  (n-p-1)!\, \d^{a_1\ldots a_{n-p-1}}_{b_1\ldots b_{n-p-1}},
\ee
leads to the commutators of the new generators
\be
\big[ K^a{}_b, E^{c_1\ldots c_p} \big] &=& (-1)^p\, p\, \d^{[c_1}_b
  E^{c_2\ldots c_p]a},\\
\big[ E^{a_1\ldots a_p}, F_{b_1\ldots b_p}\big] &=& -\frac{p}{n-1}p!
\,\d^{a_1\ldots a_p}_{b_1\ldots b_p}K + p\cdot p!\, \d^{[a_1\ldots
    a_{p-1}}_{[b_1\ldots b_{p-1}} K^{a_p]}{}_{b_p]},
\ee
and similar expressions for the magnetic generator with $p$ replaced
by $(n-p-1)$. From this, together with the commutator
(\ref{dualgravcomm}), we can deduce the following $\ell=0$
equation of motion for the $\s$-model in this approximation
\be
\label{einsteq2}
&&n\pt(n^{-1} P_{ab}) = 2 P_{c(a}Q_{b)c} +
\frac1{2(n-2)!}\bigg(\d_{ab} P_{c_1\ldots c_{n-2},c_{n-1}}P_{c_1\ldots
  c_{n-2},c_{n-1}} \nn\\&&\quad\quad-P_{c_1\ldots c_{n-2},a}P_{c_1\ldots c_{n-2},b}
-(n-2) P_{ac_1\ldots c_{n-3},c_{n-2}}P_{bc_1\ldots c_{n-3},c_{n-2}}\bigg)
\nn\\
&&\quad+\frac{p}{2\,p!} P^{\rm{el}}_{a c_1\ldots c_{p-1}}
  P^{\rm{el}}_{bc_1\ldots c_{p-1}}
- \frac{p}{2\,p!(n-1)}\d_{ab} P^{\rm{el}}_{c_1\ldots c_p}
  P^{\rm{el}}_{c_1\ldots c_p}\\
&&\quad+\frac{n-p-1}{2\,(n-p-1)!} P^{\rm{magn}}_{a c_1\ldots c_{n-p-2}}
  P^{\rm{magn}}_{bc_1\ldots c_{n-p-2}}
- \frac{p}{2\,p!(n-1)}\d_{ab} P^{\rm{magn}}_{c_1\ldots c_{n-p-1}}
  P^{\rm{magn}}_{c_1\ldots c_{n-p-1}}.\nn
\ee

By comparing (\ref{einsteq1}) and (\ref{einsteq2}) we find
agreement if we identify the new terms by
\be
P^{\rm{el}}_{a_1\ldots a_p}(t)  &\equiv& F_{ta_1\ldots a_p} (t,x_0),\\
P^{\rm{magn}}_{a_1\ldots a_{n-p-1}} (t)&\equiv& \frac{N}{(p+1)!}
\ve_{a_1\ldots a_{n-p-1}b_1\ldots b_{p+1}}F_{b_1\ldots
  b_{p+1}}(t,x_0).
\ee
in addition to (\ref{einstid}). There are sign ambiguities here, since the
fields appear quadratically in (\ref{einsteq2}). These will be fixed
from the equations of motion and Bianchi identities for the $p$-form
fields.

\end{subsection}

\end{section}

\end{document}